\documentclass[10pt]{elsart}
\usepackage{epsfig,amssymb}
\journal{Annals of Physics}

\begin{document}
\begin{frontmatter}

\title{Listening to the Universe with Gravitational-Wave Astronomy}

\author{Scott A.\ Hughes}
\address{Kavli Institute for Theoretical Physics, University of
California, Santa Barbara, CA 93106}
\address{Department of Physics, Massachusetts Institute of
Technology, Cambridge, MA 02139}
\ead{sahughes@mit.edu}

\begin{abstract}
The LIGO (Laser Interferometer Gravitational-Wave Observatory)
detectors have just completed their first science run, following many
years of planning, research, and development.  LIGO is a member of
what will be a worldwide network of gravitational-wave observatories,
with other members in Europe, Japan, and --- hopefully --- Australia.
Plans are rapidly maturing for a low frequency, space-based
gravitational-wave observatory: LISA, the Laser Interferometer Space
Antenna, to be launched around 2011.  The goal of these instruments is
to inaugurate the field of {\it gravitational-wave astronomy}: using
gravitational-waves as a means of listening to highly relativistic
dynamical processes in astrophysics.  This review discusses the
promise of this field, outlining why gravitational waves are worth
pursuing, and what they are uniquely suited to teach us about
astrophysical phenomena.  We review the current state of the field,
both theoretical and experimental, and then highlight some aspects of
gravitational-wave science that are particularly exciting (at least to
this author).
\end{abstract}
\end{frontmatter}

\section{Motivation}

The current state of gravitational-wave science is very similar to the
state of neutrino science circa 1950 {\cite{barish_ucsb}}: we have a
mature theoretical framework describing this form of radiation; we
have extremely compelling {\it indirect} evidence of the radiation's
existence; but an unambiguous direct detection has not yet happened.
Unlike the case of neutrinos, however, it is unlikely that a bright
laboratory source of gravitational radiation (analogous to the
Savannah River nuclear reactor) will be constructed (though see
{\cite{chiao}} for an alternative view).  The only guaranteed sources
of gravitational waves bright enough to be measurable will arise from
violent astrophysical events.  Though perhaps somewhat frustrating on
the one hand --- we must remain patient while we wait for nature to
supply us with a radiation source bright enough for our fledgling
detectors --- it offers a great opportunity on the other.
Gravitational radiation promises to open a unique window onto
astrophysical phenomena that may teach us much about ``dark''
processes in the universe.  Once these detectors have met their
``physics goal'' of directly and unambiguously detecting gravitational
waves, they will grow into observatories that --- we hope! --- will be
rich sources of data on violent astrophysical events.

The properties of gravitational radiation and the processes that drive
its emission are quite different from the properties and processes
relevant to electromagnetic radiation.  Consider the following
differences:

\begin{itemize}

\item Electromagnetic waves are oscillations of electric and magnetic
fields that propagate through spacetime.  Gravitational waves are
oscillations of spacetime itself.  Formally, this is an extremely
important difference, and historically has been a source of some
controversy regarding the validity of certain computation schemes in
gravitational-wave theory (with some members of the relativity
community worrying that analogies to electromagnetic radiation were
used without sufficient justification).  This difference can make it
difficult to define what exactly a gravitational wave {\it is}.  One
must identify an oscillating contribution to the curvature of
spacetime that varies on a lengthscale $\lambda/2\pi$ much shorter
than the lengthscales over which all other important curvatures vary.
In this sense, gravitational waves are more similar to waves
propagating over the ocean's surface (varying on a lengthscale much
smaller than the Earth's radius of curvature) than they are to
electromagnetic radiation.

\item Astrophysical electromagnetic radiation typically arises from
the incoherent superposition of waves produced by many emitters (e.g.,
electrons in the solar corona, hot plasma in the early universe).
This radiation directly probes the thermodynamic state of a system or
an environment.  Gravitational waves are coherent superpositions
arising from the bulk dynamics of a dense source of mass-energy.
These waves directly probe the dynamical state of a system.

\item Electromagnetic waves interact strongly with matter;
gravitational waves do not.  This follows directly from the relative
strength of the electromagnetic and gravitational interactions.  The
weak interaction strength of gravitational waves is both blessing and
curse: it means that gravitational waves propagate from emission to
observers on the Earth with essentially zero absorption, making it
possible to probe astrophysics that is hidden or dark --- e.g., the
coalescence and merger of black holes, the collapse of a stellar core,
the dynamics of the early universe.  This also means that the waves
interact very weakly with detectors, necessitating a great deal of
effort to ensure their detection.  Also, because many of the best
sources are hidden or dark, they are very poorly understood today ---
we know very little about what are likely to be some of the most
important sources of gravitational waves.

\item The direct observable of gravitational radiation is the waveform
$h$, a quantity that falls off with distance as $1/r$.  Most
electromagnetic observables {\cite{radioastronomy_note}} are some kind
of energy flux, and so fall off with a $1/r^2$ law.  This means that
relatively small improvements in the sensitivity of gravitational-wave
detectors can have a large impact on their science: doubling the
sensitivity of a detector doubles the distance to which sources can be
detected, increasing the volume of the universe to which sources are
measurable by a factor of 8.  Every factor of two improvement in the
sensitivity of a gravitational-wave observatory should increase the
number of observable sources by about an order of magnitude.

\item Electromagnetic radiation typically has a wavelength smaller
than the size of the emitting system, and so can be used to form an
image of the source, exemplified by the many beautiful images
observatories have provided over the years.  By contrast, the
wavelength of gravitational radiation is typically comparable to or
larger than the size of the radiating source.  Gravitational waves
{\it cannot} be used to form an image.  Instead, gravitational-waves
are best thought of as analogous to sound: the two polarizations carry
a stereophonic description of the source's dynamics.  Many researchers
in gravitational-wave physics illustrate their work by playing audio
encodings of expected gravitational-wave sources and of detector
noise.  Some source examples from this author's research can be found
at {\cite{sounds}}; I leave it to the reader to judge whether they are
beautiful or not.

\item In most cases, electromagnetic astronomy is based on deep
imaging of small fields of view: observers obtain a large amount of
information about sources on a small piece of the sky.
Gravitational-wave astronomy, by contrast, will be a nearly all-sky
affair: gravitational-wave detectors have nearly $4\pi$ steradian
sensitivity to events over the sky.  A consequence of this is that
their ability to localize a source on the sky is not good by usual
astronomical standards; but, it means that any source on the sky will
be detectable, not just sources towards which the detector is
``pointed''.  The contrast between the all-sky sensitivity but poor
angular resolution of gravitational-wave observatories, and the
pointed, high angular resolution of telescopes is very similar to the
angular resolution contrast of hearing and sight, strengthening the
useful analogy of gravitational waves with sound.

\end{itemize}

These differences show why we believe that gravitational-wave
astronomy will open a radically new observational window for
astrophysics, and motivate the efforts to construct sensitive
gravitational-wave detectors.  The last two points in particular
explain why we have chosen to describe gravitational-wave astronomy as
``listening to the universe''.  (Marcia Bartusiak similarly expanded
on this theme in her very engaging book ``Einstein's Unfinished
Symphony'' {\cite{marcia}}.)  Gravitational-wave astrophysics can be
thought of as learning to speak the language of gravitational-wave
sources so that we can understand and learn about the sources that the
new detectors will measure.

This article surveys the current state of this field.  Sections
{\ref{sec:concepts}} and {\ref{sec:bands}} are review material ---
Sec.\ {\ref{sec:concepts}} discusses the major background concepts
associated with gravitational radiation and gravitational-wave
detectors, and Sec.\ {\ref{sec:bands}} surveys astrophysical sources
and detection methods, categorizing them by the frequency band in
which they primarily radiate.  We then focus on several aspects of
gravitational-wave astronomy involving black holes that are of
particular interest to this author.  Section {\ref{sec:BBH}} discusses
the importance of binary black hole systems as sources of
gravitational waves, and what can be learned from such observations
from the standpoint of astrophysics and physics generally.  Section
{\ref{sec:bothrodesy}} discusses in detail a special kind of binary
black hole system --- extreme mass ratio binaries, in which one black
hole in the binary is far more massive than the other.  We discuss the
particularly powerful and interesting analyses that measurement of
these waves can make possible, and then review the challenges that
must be overcome to understand the language of these sources.

\section{Major concepts of gravitational-wave physics}
\label{sec:concepts}

The idea that radiation of some sort might be associated with the
gravitational interaction has a surprisingly long pedigree.  As early
as 1776, Laplace {\cite{laplace}} suggested that an apparent secular
acceleration in the Moon's orbit (deduced by Edmund Halley from a
study of medieval solar eclipses recorded by Al-Batanni and of still
older eclipses recorded by Ptolemy {\cite{dank_thesis}}) could be
explained by requiring that the gravitational interaction propagate at
finite speed.  (The correct explanation of this effect turned out to
be tidal transfer of the Earth's rotational angular momentum to the
Moon's orbit {\cite{dank_thesis}}.)  Poincar\'e somewhat tentatively
resurrected this idea in 1908 in an attempt to explain the anomalous
perihelion shift of Mercury {\cite{poincare}}.  (This effect was
eventually explained by the nonlinear ``post-Newtonian'' effect of
relativistic gravity {\cite{bigal0}}.)

Gravitational waves finally and (almost) unambiguously entered the
lexicon of physics as a natural consequence of general relativity.
Soon after general relativity was introduced, Einstein predicted the
existence of gravitational waves in a 1916 paper {\cite{bigal1}}.
This analysis was flawed by a few important algebraic errors, which
were corrected in a 1918 paper {\cite{bigal2}}.  Einstein showed that
gravitational radiation arises from variations in a source's
quadrupole moment, and derived (with a factor of 2 error) what has
come to be called the ``quadrupole formula'' for the rate at which the
radiation carries energy away from the source.  This is what one
expects intuitively --- gravitational waves arise from the
acceleration of masses in a manner similar to the generation of
electromagnetic radiation from the acceleration of charges.  At lowest
order, electromagnetic waves come from the time changing charge dipole
moment, and are thus dipole waves; monopole EM radiation would violate
charge conservation.  We expect (at lowest order) gravitational waves
to come from the time changing quadrupolar distribution of mass and
energy, since monopole gravitational waves would violate mass-energy
conservation, and dipole waves would violate momentum or angular
momentum conservation.

The parenthetical ``almost'' at the beginning of the preceding
paragraph refers to a rather lengthy controversy over the formal
underpinnings of gravitational radiation calculations.  These
controversies mostly came to an end in the 1980s, thanks in large part
to the careful, rigorous calculations of Thibault Damour and
collaborators (cf.\ Ref.\ {\cite{damour_prl}} and references therein)
and the excellent correspondence to observations of the Hulse-Taylor
binary pulsar {\cite{hulse_taylor,taylor_weisberg}}; see Ref.\
{\cite{dank_thesis}} for extended discussion.  It is now generally
accepted that Einstein's original quadrupole formula (corrected for
the factor of 2 error) properly describes at lowest order the energy
flow from a radiating source (even if that source has strong self
gravity, a major issue contributing to the aforementioned
controversy), and we are likewise confident that theory can go well
beyond this lowest order (see, e.g., the review by Blanchet
{\cite{blanchet_lr}} and references therein).

\begin{figure}[ht]
\includegraphics[width = 13cm]{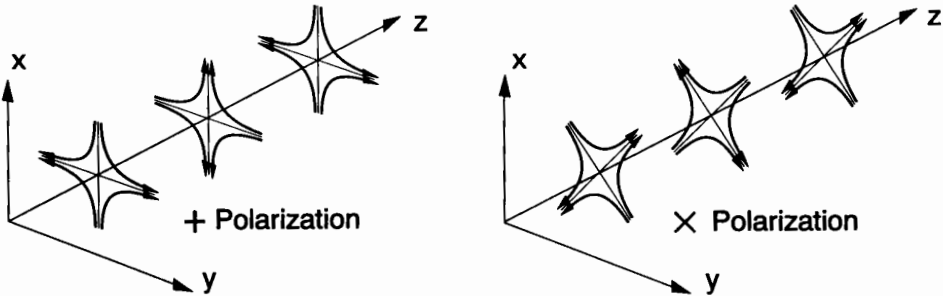}
\caption{The lines of force associated with the two polarizations of
a gravitational wave (from Ref.\ {\cite{ligoscience}}).}
\label{fig:forcelines}
\end{figure}

Gravitational waves act tidally, stretching and squeezing any object
that they pass through.  Their quadrupolar character means that they
squeeze along one axis while stretching along the other.  When the
size of the object that the wave acts upon is small compared to the
wavelength (as is the case for LIGO), forces that arise from the two
GW polarizations act as in Fig.\ {\ref{fig:forcelines}}.  The
polarizations are named ``$+$'' (plus) and ``$\times$'' (cross)
because of the orientation of the axes associated with their force
lines.

\begin{figure}[t]
\includegraphics[width = 13cm]{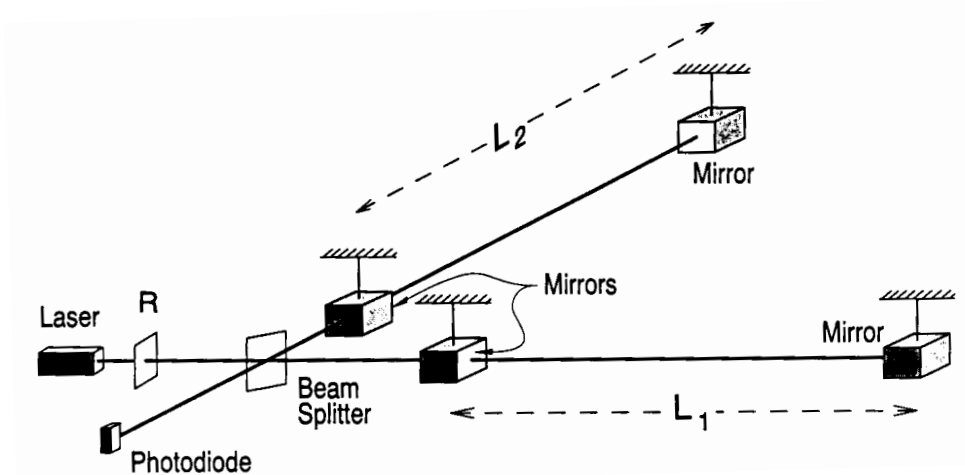}
\caption{Layout of an interferometer for detecting gravitational waves
(from Ref.\ {\cite{ligoscience}}).}
\label{fig:ifo}
\end{figure}

\begin{figure}[t]
\includegraphics[width = 13cm]{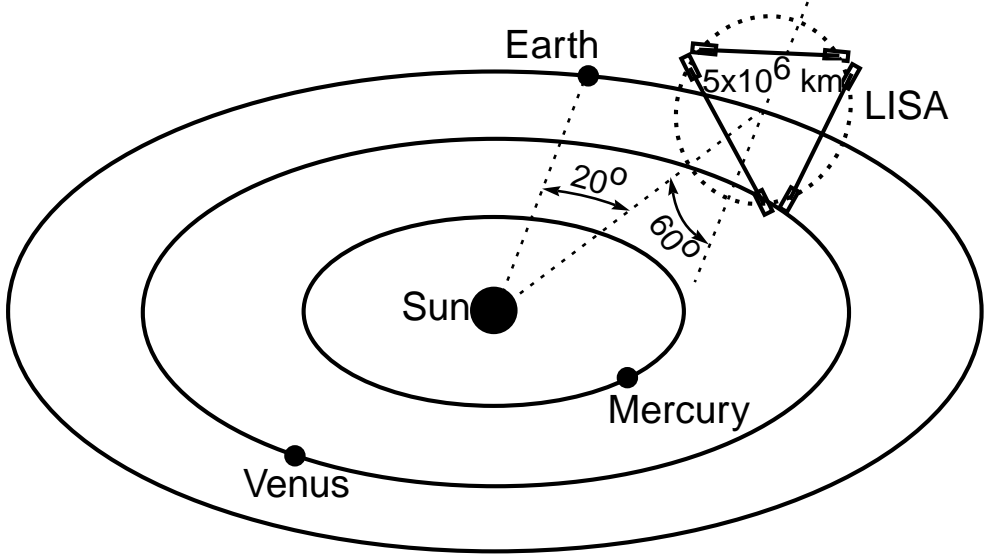}
\caption{Orbital configuration of the LISA antenna.}
\label{fig:lisa_orbit}
\end{figure}

Interferometric gravitational-wave detectors measure this tidal field
by observing their action upon a widely-separated set of test masses.
In ground-based interferometers, these masses are arranged as in Fig.\
{\ref{fig:ifo}}.  The space-based detector LISA arranges its test
masses in a large equilateral triangle that orbits the sun,
illustrated in Fig.\ {\ref{fig:lisa_orbit}}.  On the ground, each mass
is suspended with a sophisticated pendular isolation system to
eliminate the effect of local ground noise.  Above the resonant
frequency of the pendulum (typically of order $1\,{\rm Hz}$), the mass
moves freely.  (In space, the masses are actually free floating.)  In
the absence of a gravitational wave, the sides $L_1$ and $L_2$ shown
in Fig.\ {\ref{fig:ifo}} are about the same length $L$.

Suppose the interferometer in Fig.\ {\ref{fig:ifo}} is arranged such
that its arms lie along the $x$ and $y$ axes of Fig.\
{\ref{fig:forcelines}}.  Suppose further that a wave impinges on the
detector down the $z$ axis, and the axes of the $+$ polarization are
aligned with the detector.  The tidal force of this wave will stretch
one arm while squeezing the other; each arm oscillates between stretch
and squeeze as the wave itself oscillates.  The wave is thus
detectable by measuring the separation between the test masses in each
arm and watching for this oscillation.  In particular, since one arm
is always stretched while the other is squeezed, we can monitor the
difference in length of the two arms:
\begin{equation}
\delta L(t) \equiv L_1(t) - L_2(t)\;.
\label{eq:delta_L}
\end{equation}
For the case discussed above, this change in length turns out to be
the length of the arm times the $+$ polarization amplitude:
\begin{equation}
\delta L(t) = h_+(t)L\;.
\label{eq:h_simple}
\end{equation}
The gravitational wave acts as a strain in the detector; $h$ is often
referred to as the ``wave strain''.  Note that it is a dimensionless
quantity.  Equation (\ref{eq:h_simple}) is easily derived by applying
the equation of geodesic deviation to the separation of the test
masses and using a gravitational-wave tensor on a flat background
spacetime to develop the curvature tensor; see Ref.\ {\cite{300yrs}},
Sec.\ 9.2.2 for details.

We obviously do not expect astrophysical gravitational-wave sources to
align themselves in as convenient a manner as described above.
Generally, both polarizations of the wave influence the test masses:
\begin{equation}
{\delta L(t)\over L} = F^+ h_+(t) + F^\times h_\times(t) \equiv h(t)\;.
\label{eq:h_def}
\end{equation}
The antenna response functions $F^+$ and $F^\times$ weight the two
polarizations in a quadrupolar manner as a function of a source's
position and orientation relative to the detector; see
{\cite{300yrs}}, Eqs.\ (104a,b) and associated text.

The energy flux carried by gravitational waves scales as $\dot h^2$
(where the overdot denotes a time derivative).  In order for the
energy flowing through large spheres to be conserved, $h$ must fall
off with distance as $1/r$.  As discussed above, the lowest order
contribution to the waves arises from changes in a source's quadrupole
moment.  To order of magnitude, this moment is given by $Q \sim
(\mbox{source mass})(\mbox{source size})^2$.  By dimensional analysis,
we then know that the wave strain must have the form
\begin{equation}
h \sim {G\over c^4}{\ddot Q\over r}\;.
\label{eq:h_ordermag}
\end{equation}
The second time derivative of the quadrupole moment is given
approximately by $\ddot Q\simeq 2M v^2\simeq 4 E^{\rm ns}_{\rm kin}$;
$v$ is the source's internal velocity, and $E^{\rm ns}_{\rm kin}$ is
the nonspherical part of its internal kinetic energy.  Strong sources
of gravitational radiation are sources that have strong non-spherical
dynamics --- for example, compact binaries (containing white dwarfs,
neutron stars, and black holes), mass motions in neutron stars and
collapsing stellar cores, the dynamics of the early universe.

Violent events that are likely to be interesting gravitational-wave
sources are very rare --- for example, supernovae from the collapse of
massive stellar cores appear to occur in our galaxy once every few
centuries.  For our detectors to have a realistic chance of measuring
observable events, they must be sensitive to sources at rather large
distances.  For example, to have an interesting shot at measuring the
coalescence of binary neutron star systems, we need to reach out to
several hundred megaparsecs ({\it i.e.}, a substantial fraction of
$10^9$ light years) {\cite{nps,phinney91,kl2000}}.  For such
coalescences, $E^{\rm ns}_{\rm kin}/c^2 \sim\mbox{1 solar
mass}\,(\equiv 1\,M_\odot)$.  Plugging into Eq.\ (\ref{eq:h_ordermag})
gives the estimate
\begin{equation}
h \sim 10^{-21} - 10^{-22}\;.
\label{eq:h_num_est}
\end{equation}
This sets the sensitivity required to measure gravitational waves.
Combining this scale with Eq.\ (\ref{eq:h_def}) tells us that for
every kilometer of baseline $L$ we need to be able to measure a
distance shift $\delta L$ of better than $10^{-16}$ centimeters.

This is usually the point at which people decide that
gravitational-wave scientists aren't playing with a full deck.  How
can we possibly hope to measure an effect that is $\sim10^{12}$ times
smaller than the wavelength of visible light?  For that matter, how is
it possible that thermal motions do not wash out such a tiny effect?

That such measurement is possible with laser interferometry was
analyzed thoroughly and published by Rainer Weiss in 1972
{\cite{weiss72}}.  (It should be noted that the possibility of
detecting gravitational waves with laser interferometers has an even
longer history, reaching back to Pirani in 1956 {\cite{pirani}}, and
has been independently invented by Gertsenshtein and Pustovoit in 1962
{\cite{gp1962}} and Weber in the 1960s (unpublished), prior to Weiss's
detailed analysis.  See Sec.\ 9.5.3 of Ref.\ {\cite{300yrs}} for
further discussion.)  Examine first how a laser with a wavelength of 1
micron can measure a $10^{-16}$ cm displacement.  In a laser
interferometer like LIGO, the basic optical layout is as sketched in
Fig.\ {\ref{fig:ifo}}.  A carefully prepared laser state is split at
the beamsplitter and sent into the Fabry-Perot arm cavities of the
detector.  The reflectivities of the mirrors in these cavities are
chosen such that the light bounces roughly 100 times before exiting
the arm cavity (that is, the finesse ${\mathcal F}$ of the cavity is
roughly 100).  This corresponds to about half a cycle of a 100 Hz
gravitational wave.  The phase shift acquired by the light during
those 100 round trips is
\begin{equation}
\Delta\Phi_{\rm GW}\sim100\times2\times\Delta L\times2\pi/\lambda
\sim 10^{-9}\;.
\label{eq:phase_estimate}
\end{equation}
This phase shift can be measured provided that the shot noise at the
photodiode, $\Delta\Phi_{\rm shot}\sim 1/\sqrt{N}$, is less than
$\Delta\Phi_{\rm GW}$.  $N$ is the number of photons accumulated over
the measurement; $1/\sqrt{N}$ is the phase fluctuation in a quantum
mechanical coherent state that describes a laser.  We therefore must
accumulate $\sim 10^{18}$ photons over the roughly $0.01$ second
measurement, translating to a laser power of about 100 watts.  In
fact, as was pointed out by Ronald Drever {\cite{drever}}, one can use
a much less powerful laser: even in the presence of a gravitational
wave, only a tiny portion of the light that comes out of the
interferometer's arms goes to the photodiode.  The vast majority of
the laser power is sent back to the laser.  An appropriately placed
mirror bounces this light back into the arms, {\it recycling} the
light.  The recycling mirror is shown in Fig.\ {\ref{fig:ifo}},
labeled ``R''.  With it, a laser of $\sim 10$ watts drives several
hundred watts of input to the interferometer's arms.

Thermal excitations are overcome by averaging over many many
vibrations.  For example, the atoms on the surface of the
interferometers' test mass mirrors oscillate with an amplitude
\begin{equation}
\delta l_{\rm atom} = \sqrt{k T\over m\omega^2}
\sim 10^{-10}\,{\rm cm}
\end{equation}
at room temperature $T$, with $m$ the atomic mass, and with a
vibrational frequency $\omega\sim10^{14}\,{\rm s}^{-1}$.  This
amplitude is huge relative to the effect of gravitational radiation
--- how can we possibly hope to measure the wave?  The answer is that
atomic vibrations are random and incoherent.  The $\sim 7$ cm wide
laser beam averages over about $10^{17}$ atoms and at least $10^{11}$
vibrations per atom in a typical measurement.  The effect is thus
suppressed by a factor $\sim\sqrt{10^{28}}$ --- atomic vibrations are
{\it completely} irrelevant compared to the coherent effect of a
gravitational wave.  Other thermal vibrations, however, are not
irrelevant and in fact dominate LIGO's noise in certain frequency
bands.  For example, the test masses' normal modes are thermally
excited.  The typical frequency of these modes is $\omega\sim
10^5\,{\rm s}^{-1}$ and they have mass $m \sim 10\,{\rm kg}$, so
$\delta l_{\rm mass} \sim 10^{-14}\,{\rm cm}$.  This, again, is much
larger than the effect we wish to observe.  However, the modes are
very high frequency, and so can be averaged away provided the test
mass is made from material with a very high quality factor $Q$ --- the
mode's energy is confined to frequencies near $\omega$ and doesn't
leak into the band we want to use for measurements.  Understanding the
physical nature of noise in gravitational-wave detectors is an active
field of current research; see Refs.\
{\cite{levin,lt2000,sl2001,bc1,bc2,ht1998,tw1999,tcreighton}} and
references therein for a glimpse of recent work.  In all cases, the
fundamental fact to keep in mind is that a gravitational wave acts
{\it coherently}, whereas noise acts {\it incoherently}, and thus can
be beaten provided one is able to average away the incoherent noise
sources.

\vfill
\section{Gravitational-wave frequency bands and measurement}
\label{sec:bands}

It is useful to categorize gravitational-wave sources (and the methods
for detecting their waves) by the frequency band in which they
radiate.  Broadly speaking, we may break the gravitational-wave
spectrum into four rather different bands: the {\it ultra low
frequency} band, $10^{-18}\,{\rm Hz}\lesssim f\lesssim 10^{-13}\,{\rm
Hz}$; the {\it very low frequency} band, $10^{-9}\,{\rm Hz}\lesssim
f\lesssim 10^{-7}\,{\rm Hz}$; the {\it low frequency} band,
$10^{-5}\,{\rm Hz}\lesssim f\lesssim 1\,{\rm Hz}$; and the {\it high
frequency} band, $1\,{\rm Hz}\lesssim f\lesssim 10^4\,{\rm Hz}$.

For compact sources (mass/energy configurations that are of compact
support), the band in which gravitational waves are generated is
typically related to the source's size $R$ and mass $M$.  $R$ is meant
to set the scale over which the source's dynamics vary; for example,
it could be the actual size of a particular body, or the separation of
members of a binary.  The ``natural'' gravitational-wave frequency of
such a source is $f_{\rm GW} \sim (1/2\pi)\sqrt{G M/R^3}$.  Because
$R\lesssim 2 G M/c^2$ (the Schwarzschild radius of a mass $M$), we can
estimate an upper bound for the frequency of a compact source:
\begin{equation}
f_{\rm GW}(M) < {1\over4\sqrt{2}\pi}{c^3\over G M} \simeq 10^4\,{\rm
Hz} \left({M_\odot\over M}\right)\;.
\end{equation}
This is a rather hard upper limit, since many interesting sources are
quite a bit larger than $2 G M/c^2$, or else evolve through a range of
sizes before terminating their emission at $R \sim 2 GM/c^2$.
Nonetheless, this frequency gives some sense of the types of compact
sources that are likely to be important in each band --- high
frequency compact sources are of stellar mass (several solar masses);
low frequency compact sources are of thousands to millions of solar
masses, or else contain widely separated stellar mass bodies; etc.
Other interesting sources of waves, particularly in the lower
frequency bands, are not well-described by these compact body rules;
we will discuss them separately in greater depth below.

\subsection{High frequency}
\label{subsec:high}

The high frequency band, $1\,{\rm Hz}\lesssim f \lesssim 10^4\,{\rm
Hz}$, is the band targeted by the new generation of ground-based laser
interferometric detectors, such as LIGO.  (It also corresponds roughly
to the audio band of the human ear: when converted to sound, LIGO
sources are human audible without any frequency scaling.)  The low
frequency end of this band is set by the fact that it is extremely
difficult to isolate against ground vibrations at low frequencies, and
probably impossible to isolate against gravitational coupling to
ground vibrations, human activity, and atmospheric motions
{\cite{ht1998,tw1999,tcreighton}}.  The high end of the band is set by
the fact that it is unlikely any interesting gravitational-wave source
radiates at frequencies higher than a few kilohertz --- from the
arguments sketched above, such a source would have to be relatively
low mass but extremely compact.

\begin{figure}[t]
\includegraphics[width = 13cm]{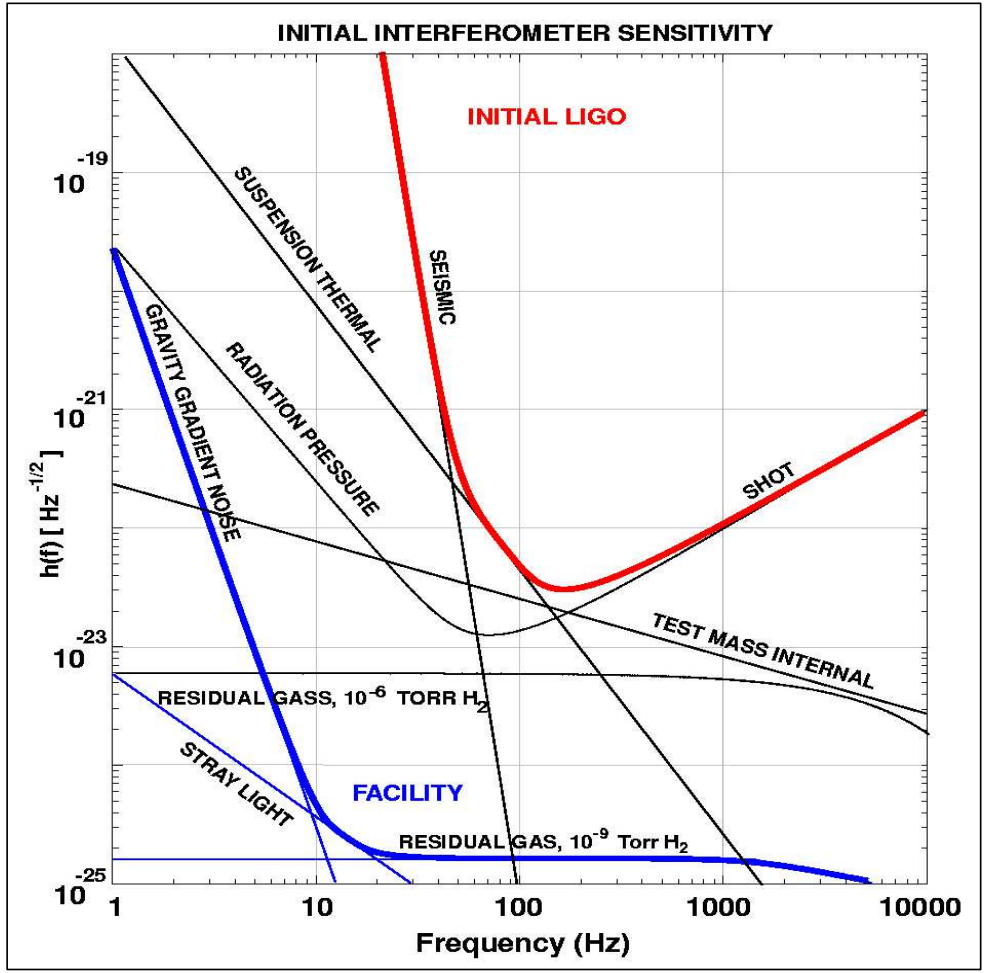}
\caption{Sensitivity goals of the initial LIGO interferometers, and
facility limits on the LIGO sensitivity (taken from Ref.\
{\cite{snowmass}}).}
\label{fig:ligo_sens}
\end{figure}

The operating principles of a ground-based laser interferometric
detector have already been sketched in Sec.\ {\ref{sec:concepts}}
[cf.\ the text following Eq.\ (\ref{eq:h_num_est})].  The curve
describing the sensitivity of such detectors typically takes a shape
similar to that shown in Fig.\ {\ref{fig:ligo_sens}}.  At high
frequencies, the detectors' sensitivities rapidly degrade because of
photon shot noise --- fluctuations in the number of photons used in
the measurement process.  Making a measurement at a frequency $f$
essentially means averaging for a timescale $T = 1/f$.  As the time
$T$ becomes shorter, a smaller number of photons are gathered in the
course of the measurement, and hence the typical fluctuation in the
number of photons is relatively more important.  At intermediate
frequencies, thermally excited normal modes in the test mass mirrors
(at the ends of the arms in Fig.\ {\ref{fig:ifo}}) and in the mirrors'
suspensions dominate the noise budget.  The resonant frequencies of
these modes are carefully chosen to be rather far above the band of
greatest interest for gravitational-wave observation; and, the $Q$ of
the masses and suspensions are made as large as is practical so that
the modes' energy bleeds into the gravitational-wave band as little as
possible.  Some contamination is of course inevitable.  At very low
frequencies, seismic motions dominate the detectors' noise.  The test
masses are carefully suspended on multi-level pendular systems to
isolate them from local ground motions.  This makes the masses
effectively free falling above the resonant frequency of the pendulum;
below that frequency, however, the noise due to ground motion
dominates the motion spectrum of the masses.

Several interferometric gravitational-wave observatories are either
operating or being completed in the United States, Europe, Japan, and
Australia.  Multiple observatories widely scattered over the globe are
extremely important, both as checks on one another for assured
detection and to aid in the interpretation of measurements.  For
example, position determination and thence measurement of the distance
to a source follows from triangulation of time-of-flight differences
between separated detectors.  The major interferometer projects are:

\begin{itemize}

\item {\bf LIGO.}  The Laser Interferometer Gravitational-Wave
Observatory currently consists of three operating interferometers: a
single four kilometer interferometer in Livingston, Louisiana, as well
as a pair of interferometers (four kilometers and two kilometers) in
the LIGO facilities at Hanford, Washington.  The sites are separated
by 3000 kilometers, and are situated to support coincidence analysis
of events.

\item {\bf Virgo.} Virgo is a three kilometer French-Italian detector
under construction near Pisa, Italy {\cite{virgo}}.  In most respects,
Virgo is quite similar to LIGO.  A major difference is that Virgo
employs a very sophisticated seismic isolation system that promises
extremely good low frequency sensitivity.

\item {\bf GEO600.} GEO600 is a six hundred meter interferometer
constructed by a German-English collaboration near Hannover, Germany
{\cite{geo}}.  Despite its shorter arms, GEO600 is expected to achieve
sensitivity comparable to the multi-kilometer instruments by
incorporating advanced interferometry techniques from the beginning.
This will make it an invaluable testbed for technology to be used in
later generations of the larger instruments, as well as enabling it to
make astrophysically interesting measurements.

\item {\bf TAMA300.} TAMA300 is a three hundred meter interferometer
operating near Tokyo.  It has been in operation for several years now
{\cite{tama}}; the most recent run achieved a displacement sensitivity
$10^{-16}\,{\rm cm}/\sqrt{\rm Hz}$ {\cite{noisecomment}} at
frequencies near 1000 Hz.  The TAMA team is currently designing a
three kilometer interferometer {\cite{advtama}}, building on their
experiences with the three hundred meter instrument.

\item {\bf ACIGA.} The Australian Consortium for Interferometric
Gravitational-Wave Astronomy is currently constructing an eighty meter
research interferometer near Perth, Australia {\cite{aciga}}, hoping
that it will be possible to extend it to multi-kilometer scale in the
future.  Such a detector would likely be a particularly valuable
addition to the worldwide stable of detectors, since all the Northern
Hemisphere detectors lie very nearly on a common plane.  An Australian
detector would be far outside this plane, allowing it to play an
important role in determining the location of sources on the sky.

\end{itemize}

All of these detectors have or will have sensitivities similar to that
illustrated in Fig.\ {\ref{fig:ligo_sens}} (which shows, in
particular, the sensitivity goal of the first generation of LIGO
interferometers).  This figure also shows the ``facility limits'' ---
the lowest noise levels that can be achieved even in principle within
an interferometer facility.  The low level facility limits come from
{\it gravity-gradient noise}: noise arising from gravitational
coupling to fluctuations in the local mass distribution (such as from
seismic motions in the earth near the test masses {\cite{ht1998}},
human activity near the detector {\cite{tw1999}}, and density
fluctuations in the atmosphere {\cite{tcreighton}}).  At higher
frequencies, the facility limit arises from residual gas (mostly
hydrogen) in the interferometer vacuum system.  Stray molecules of gas
effectively cause stochastic fluctuations in the index of refraction,
a source of noise as we try to make ever more precise measurements.

There's a great deal of room for improvement between the sensitivity
goals of the first detectors and the facility limits.  Much active
research and development work is geared towards developing improved
interferometers which will have greater astrophysical reach than the
first generation of detectors.  The first detectors have been designed
somewhat conservatively, ensuring that they can be operated for
several years without requiring too much technology development.
Upgraded detectors will have the seismic ``wall'' pushed down to lower
frequencies and will have noise curves that are moderately
``tunable'', shaping the detector response to chase down signals that
are particularly interesting or important
{\cite{bc1,bc2,whitepaper,tuning}}.  We should emphasize that, at
present, much effort is being put into reaching the initial
sensitivity goals.  The LIGO detectors have made enormous strides in
improving their sensitivity recently (gaining several orders of
magnitude over the course of 2002), but are still some distance from
the design goals.  Seismic noise in particular has proven to be a
greater problem than was anticipated (largely because of increased
human activity near the two LIGO sites), so improvements to the test
masses' isolation systems will be implemented quite quickly.

In the remainder of this subsection, we take a quick tour of some of
the more well-understood possible sources of measurable gravitational
waves in the high-frequency band.  We emphasize at this point that
such a listing of sources can in no way be considered comprehensive:
we are hopeful that some gravitational-wave sources may surprise us,
as has been the case whenever we have studied the universe with a new
type of radiation.  If we regard gravitational-wave astrophysics as
learning to speak the language of gravitational-wave sources, then
surprise sources will be somewhat akin to discovering a lost language
written in an unknown script --- interpreting and understanding their
message will be quite difficult.

\subsubsection{Compact binaries} 

Compact binaries --- binary star systems in which each member is a
collapsed, compact stellar corpse (neutron star or black hole) --- are
currently the best understood sources of gravitational waves.  Double
neutron stars have been studied observationally since the mid 1970s;
three such systems {\cite{phinney91}} tight enough to merge within a
few $10^8$ or $10^9$ years have been identified in the galaxy (two in
the galactic field, one in a globular cluster).  Detailed studies of
these systems currently provide our best data on gravitational-wave
generation {\cite{hulsetaylor,taylorweisberg,stairs}}, and led to the
1993 Nobel Prize for Joseph Taylor and Russell Hulse.  Extrapolation
from these observed binaries in the Milky Way to the universe at large
{\cite{nps,phinney91,kl2000}} indicates that gravitational-wave
detectors should measure at least several and at most several hundred
binary neutron star mergers each year (following detector upgrades;
the rates for initial detectors suggest that detection is plausible
but not very probable --- the expected rate is of order one per
decade).  Population synthesis (modeled evolution of stellar
populations) indicates that the measured rate of binaries containing
black holes should likewise be interestingly large (perhaps even for
initial detectors)
{\cite{bb98,pzy98,fwh99,pzw2000,bkb2002,nstt97,ictn98}}.  The
uncertainties of population synthesis calculations are rather large,
however, due to poorly understood aspects of stellar evolution and
compact binary formation; data from gravitational-wave detectors is
likely to have a large impact on this field.

We will revisit and discuss in greater depth this class of sources in
Sec.\ {\ref{sec:BBH}}.

\subsubsection{Stellar core collapse}

Core collapse in massive stars (the engine of Type II supernova
explosions) has long been regarded as likely to be an important source
of gravitational waves; see, for example, Ref.\ {\cite{eardley83}} for
an early review.  Stellar collapse certainly exhibits all of the {\it
necessary} conditions for strong gravitational-wave generation: large
amounts of mass ($1 - 100\,M_\odot$) flow in a compact region
(hundreds to thousands of kilometers) at relativistic speeds ($v/c
\sim 1/5$).  However, these conditions are not {\it sufficient} to
guarantee strong emission.  In particular, the degree of asymmetry in
collapse is not particularly well understood [cf.\ the text following
Eq.\ (\ref{eq:h_ordermag}), arguing that non-spherical dynamics drives
gravitational-wave emission].  If stellar cores are rapidly rotating,
instabilities can develop that are certain to drive strong
gravitational-wave emission.  An example of such an instability is the
development of a rapidly rotating bar-like mode in the dense material
of the stellar core {\cite{chandra69,cent2000,new2000}}.  Such an
instability has a rapidly varying quadrupole moment and potentially
generates copious amounts of gravitational waves.

Fryer, Holz, and Hughes {\cite{fhh}} recently surveyed the status of
core-collapse simulations with an eye to understanding whether such
collapses are likely to produce interesting and measurable waves.
They find that stellar cores in fact are quite likely to have enough
angular momentum to be susceptible to secular or dynamical
instabilities such as the bar mode. The detectability of the waves
from these modes will depend quite strongly on the coherence of the
emission mechanism: detectable waves arise from modes that hold
together long enough to radiate several tens of gravitational-wave
cycles without changing their peak frequency too strongly.  Even in
this case, observers will need to wait for upgrades before such
detection is likely to become commonplace (unless we get lucky and a
star collapses relatively close by).  Future theoretical progress in
this field will come from detailed three-dimensional simulations of
core collapse processes.  We note that significant progress has been
made on this problem recently {\cite{fw2002}}, and are confident that
we will have a grasp of core collapse wave emission robust enough to
enable the design of useful detection algorithms and astrophysical
studies by the time that the upgraded detectors are likely to be
operating.

\subsubsection{Periodic emitters}

Periodic sources of gravitational waves radiate at constant or nearly
constant frequency, like radio pulsars.  In fact, the prototypical
source of continuous gravitational waves is a rotating neutron star,
or gravitational-wave pulsar.  A non-axisymmetric neutron star
(caused, for example, by a crust that is somewhat oblate and
misaligned with the star's spin axis) will radiate gravitational waves
with characteristic amplitude
\begin{equation}
h_c \sim {G\over c^4}{I f^2 \epsilon \over r}\;,
\end{equation}
where $I$ is the star's moment of inertia, $f$ is the wave frequency,
and $r$ is the distance to the source.  The crucial parameter
$\epsilon$ characterizes the degree to which the star is distorted; it
is rather poorly understood.  Various mechanisms have been proposed to
explain how a neutron star can be distorted to give a value of
$\epsilon$ interesting as a gravitational-wave source; see
{\cite{jones,ben}} for further discussion.  Examples of some
interesting mechanisms include misalignment of a star's internal
magnetic field with the rotation axis {\cite{cutler02}} and distortion
by accreting material from a companion star {\cite{lars,ucb}}.

Whatever the mechanism generating the distortion, it is clear that
$\epsilon$ will be relatively small, so that $h_c \sim 10^{-24}$ or
smaller --- rather weak.  (Note that if these sources were not weak
emitters, the backreaction of gravitational-wave emission would make
their frequencies change more quickly --- they would not be periodic
emitters.)  Measuring these waves will require coherently tracking
their signal for a large number of wave cycles --- coherently tracking
$N$ cycles boosts the signal strength by a factor $\sim\sqrt{N}$.
This is actually fairly difficult, since the signal is strongly
modulated by the Earth's rotation and orbital motion, ``smearing'' the
waves' power across multiple frequency bands.  Searching for periodic
gravitational waves means demodulating the motion of the detector, a
computationally intensive problem since the modulation is different
for every sky position.  Unless one knows in advance the position of
the source, one needs to search over a huge number of sky position
``error boxes'', perhaps as many as $10^{14}$.  One rapidly becomes
computationally limited.  (Note that radio pulsar searches face this
same problem, with the additional complication that radio pulses are
dispersed by the interstellar medium.  However, in this case, it is
known in advance which sky position is being examined, so the
computational cost is usually not as great.)  For further discussion,
see {\cite{periodic}}; for ideas about doing hierarchical searches
that require less computer power, see {\cite{per_hier}}.

Finally, we note that the r-mode instability (a source of waves from a
current instability in rotating neutron stars) would generate waves
that are nearly periodic {\cite{andersson,fm98,lom98,nks99,olcsva98}}.
Although the physics of this source is rather different from the
physics of bumpy neutron stars, the character of the waves is quite
similar, at least as far as detection goes.  We note, though, that
recent results {\cite{rls2000,lo2002,arrasetal,gressmanetal}} indicate
that the r-mode is suppressed rather more robustly than previously
appreciated.  Conventional wisdom currently suggests that r-mode waves
are unlikely to be important sources from isolated neutron stars,
though r-modes driven by accretion from a companion may turn out to be
quite important {\cite{heyl}}.  See {\cite{fa2002}} for further
discussion.

\subsubsection{Stochastic backgrounds}

Stochastic backgrounds are ``random'' gravitational waves, arising
from a large number of independent, uncorrelated sources that are not
individually resolvable.  A particularly interesting source of
backgrounds is the dynamics of the early universe --- an all-sky
gravitational-wave background, similar to the cosmic microwave
background.  Backgrounds can arise from amplification of primordial
fluctuations in the universe's geometry, phase transitions as
previously unified interactions separated, or the condensation of a
brane from a higher dimensional space.  These waves can actually
spread over a wide range of frequency bands; waves from inflation in
particular span all bands, from ultra low frequency to high frequency.
We will discuss such inflationary waves in greater detail in Sec.\
{\ref{subsec:ultralow}}; here, we briefly discuss how these
backgrounds are characterized at higher frequencies, and the
sensitivity to them that LIGO should achieve.

Stochastic backgrounds are described by their contribution to the
universe's energy density, $\rho_{\rm gw}$.  In particular, one is
interested in the energy density as a fraction of that needed to close
the universe, over some frequency band:
\begin{equation}
\Omega_{\rm gw}(f) = {1\over\rho_{\rm crit}}{d\rho_{\rm gw}\over
d\ln f}\;,
\label{eq:stoch_gw_om}
\end{equation}
where $\rho_{\rm crit} = 3 H_0^2/8\pi G$ is the critical density
needed to close the universe.  ($H_0$ is the value of the Hubble
constant today.)  Different cosmological sources produce different
levels of $\Omega_{\rm gw}(f)$, centered in different bands.  In the
high frequency band, waves produced by inflation are likely to be
rather weak: estimates suggest that the spectrum will be flat across
LIGO's band, with magnitude $\Omega_{\rm gw} \sim 10^{-15}$ at best
{\cite{turner1}}.  Waves from phase transitions can be significantly
stronger, but are typically peaked around a frequency that depends on
the temperature $T$ of the phase transition {\cite{bruce96,kmk2001}}:
\begin{equation}
f_{\rm peak} \sim 100\,{\rm Hz}\left({T\over10^5\,{\rm TeV}}\right)\;.
\label{eq:freq_phasetrans}
\end{equation}
The temperature required to enter the LISA band, $f\sim 10^{-4} -
10^{-2}$ Hz, is $T \sim 100 - 1000\,{\rm GeV}$, nicely corresponding
to the electroweak phase transition.  Waves arising from
extradimensional dynamics should peak at a frequency given by the
scale $b$ of the extra dimensions {\cite{hogan_prl,hogan_prd}}:
\begin{equation}
f_{\rm peak} \sim 10^{-4}\,{\rm Hz}\left({1\,{\rm mm}\over
b}\right)^{1/2}\;.
\label{eq:freq_extradim}
\end{equation}
For the waves to be in LIGO's band, the extra dimensions must be
rather small, $b \sim 10^{-15}$ meters.  LISA's band is accessible for
a scale similar to those discussed in modern brane-world work
{\cite{rs1,rs2}}.  It's worth noting that extradimensional models
which attempt to explain the acceleration of the universe typically
predict relic spectra of gravitational waves that are rather large,
and thus may be falsified by gravitational-wave observations
{\cite{sss2002}}.

Because of their random nature, stochastic gravitational waves look
just like noise.  Ground-based detectors will measure stochastic
backgrounds by comparing data at multiple sites and looking for
``noise'' that is correlated {\cite{maggiore,ar}}.  For comparing to a
detector's noise, one should construct the characteristic stochastic
wave strain,
\begin{equation}
h_c \propto f^{-3/2} \sqrt{\Omega_{\rm gw}(f) \Delta f}\;.
\end{equation}
(For further discussion and the proportionality constants, see
{\cite{maggiore}}.)  Note that this strain level grows sharply with
decreasing frequency.  As we will discuss in Sec.\
{\ref{subsec:verylow}}, observations in the very low frequency band
are likely to provide the best constraints on stochastic waves in the
near future.

Early LIGO detectors will have fairly poor sensitivity to the
background, constraining it to a level $\Omega_{\rm gw}\sim
5\times10^{-6}$ in a band from about 100 Hz to 1000 Hz.  This is
barely more sensitive than known limits from cosmic nucleosynthesis
{\cite{bruce96}}.  Later upgrades will be significantly more
sensitive, able to detect waves with $\Omega_{\rm gw}\sim 10^{-10}$,
which is good enough to place interesting limits on cosmological
backgrounds.

\subsection{Low frequency}
\label{subsec:low}

There is no hope of measuring gravitational waves in the low frequency
band, $10^{-5}\,{\rm Hz}\lesssim f\lesssim 1\,{\rm Hz}$, using a
ground-based instrument: even if it were possible to completely
isolate one's instrument from local ground motions, gravitational
coupling to fluctuations in the local mass distribution ultimately
limits the sensitivity to frequencies $f \gtrsim 1\,{\rm Hz}$.  As we
shall discuss below, however, many extremely interesting
gravitational-wave sources radiate in this band.  The only way to
measure these waves is to build a gravitational-wave observatory in
the quiet environment of space, far removed from low-frequency noise
sources.

Such an instrument is currently being designed jointly by NASA in the
United States and ESA, the European Space Agency: LISA, the Laser
Interferometer Space Antenna.  If all goes well, LISA would be
launched into orbit in or near 2011.  Like LIGO, LISA will be a laser
interferometer --- changes in the distance between widely separated
test masses will be monitored to find variations consistent with the
action of gravitational waves.  However, the scale of LISA is vastly
different from that of LIGO, and so details of its operations are
quite different.  In particular, LISA has armlengths $L \simeq 5
\times 10^6\,{\rm km}$, vastly larger than LIGO and all other
ground-based detectors.  The three spacecraft which delineate the ends
of LISA's arms are placed into orbits such that LISA forms a
triangular constellation orbiting the sun, inclined $60^\circ$ with
respect to the plane of the ecliptic and following the Earth with a
$20^\circ$ lag.  This configuration is sketched in Fig.\
{\ref{fig:lisa_orbit}}.  Since it essentially shares Earth's orbit,
the constellation orbits the sun once per year, ``rolling'' as it does
so.  This orbital motion plays an important role in pinpointing the
position of gravitational-wave sources by modulating the measured
waveform --- the modulation encodes source location and makes position
determination possible.

Each spacecraft contains two optical assemblies, each of which houses
a 1 watt laser and a 30 centimeter telescope.  Because of the extreme
lengths of the interferometer's arms, Fabry-Perot interferometry as in
LIGO is not at all possible: diffraction spreads the laser beam over a
diameter of about $20\,{\rm km}$ as it propagates the $5 \times
10^6\,{\rm km}$ from one spacecraft to the other.  With this much
spread, multiple bounces in LISA's arms obviously aren't feasible.
Instead, a portion of that $20\,{\rm km}$ wavefront is sampled with
the telescope.  That light is then interfered with a sample of light
from the on-board laser.  Each spacecraft thus generates two
interference data streams; six signals are generated by the full LISA
constellation.  From these six signals, we can construct the time
variations of LISA's armlengths and then build both gravitational-wave
polarizations.  More information and details can be found in Refs.\
{\cite{prephaseA,lisa1,lisa2,jointproposal}}.

It is worth noting at this point that the LISA armlengths are {\it
not} constant --- as the constellation orbits, the distances between
the various spacecraft vary by about $1\%$ (including effects such as
planetary perturbations).  This is {\it far} larger than the effect
produced by gravitational waves, which is of order picometers.
However, these variations occur over timescales of order months, and
are extremely smooth and well modeled.  It will not be difficult to
fit out these very low frequency variations, leaving clean data in the
interesting low-frequency gravitational-wave band.  Note also that
these picometer scale variations are not too difficult to measure in
this frequency band: measuring in this band entails gathering photons
for a time $10\,{\rm sec} \lesssim T \lesssim 1\,{\rm day}$.  Even
though the bulk of the laser's emitted power is lost due to
diffraction, enough photons are gathered on this timescale that the
phase shift due to the gravitational-wave can be determined [cf.\ the
argument outlined in and near Eq.\ (\ref{eq:phase_estimate})].

The gravitational-wave signals are actually read out by monitoring the
position of the so-called ``gravitational sensor'' on each optical
assembly; in particular, the position of a ``proof mass'' which floats
freely and constitutes the test mass for the LISA antenna is
monitored.  Because it is freely floating, the proof mass responds
solely to gravitational forces (or, in relativistic language, follows
a geodesic of the spacetime).  Micronewton thrusters keep the bulk
spacecraft centered on these proof masses, forcing the craft to follow
the average trajectory of the two proof masses.  In this way, LISA is
isolated from low frequency noises that could impact the ability to
measure gravitational waves (e.g., variations in solar radiation
pressure).  This is called a drag-free system, since such systems were
first used to reduce the effect of Earth's atmospheric drag on low
altitude satellites.

\begin{figure}[t]
\includegraphics[width = 13cm]{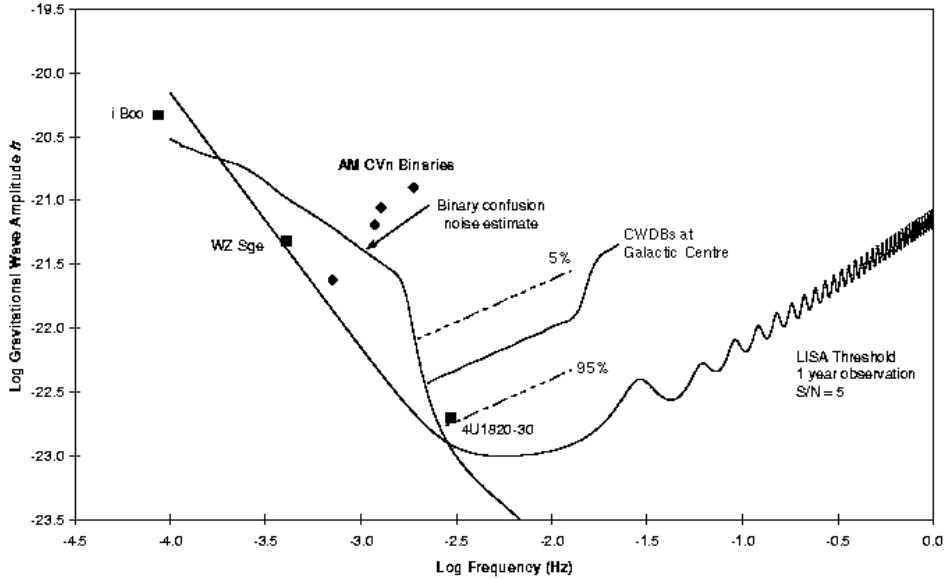}
\caption{LISA sensitivity, including a few interesting known sources,
taken from Ref.\ {\cite{snowmass}}.  Points are the expected signal
amplitude of certain known monochromatic binary stars.  ``CWDB''
stands for close white dwarf binary.}
\label{fig:lisa_sens}
\end{figure}

The sensitivity of LISA to gravitational waves is shown in Fig.\
{\ref{fig:lisa_sens}}.  At high frequencies, the noise budget is
dominated by the accuracy with which laser interferometry can
determine variations in the $5\times 10^6\,{\rm km}$ distance between
proof masses on distant spacecraft, which is largely limited by photon
shot noise.  Wiggles in the sensitivity curve at this point arise
because, in this band, the gravitational wavelength is shorter than
LISA's armlength; see {\cite{lhh}} for further discussion.  At lower
frequencies, the instrumental noise is dominated by spurious
accelerations on the proof mass.  LISA requires that these
accelerations be kept at a level below $3\times 10^{-15}\,{\rm
m/sec}^2\,{\rm Hz}^{-1/2}$ in this band.  This subsystem will be
tested by SMART-2 (Small Mission for Advanced Research and
Technology), to be launched in 2006 by ESA with participation from
NASA.

Note in Fig.\ {\ref{fig:lisa_sens}} the curve labeled ``Binary
confusion estimate'' over the band $10^{-4}\,{\rm Hz}\lesssim f
\lesssim 3\times10^{-3}\,{\rm Hz}$.  In this band, LISA's ``noise''
actually comes not from the instrument itself but from a confused
stochastic background of gravitational waves!  It is expected that so
many binary star systems (primarily double white dwarf binaries) in
the galaxy will be radiating in this band that we will not have
sufficient information to resolve them --- $10^2 - 10^4$ binaries may
contribute to the waves measured in a single frequency bin of width
$\delta f \sim 10^{-7}\,{\rm Hz}$ {\cite{cutler98}}.  This confused
background of waves is ``noise'' from the point of view of observers
wishing to measure other sources in this band (though of course it is
extremely interesting ``signal'' to an astrophysicist interested in
close binary populations).

This aspect of LISA's ``noise'' budget points to an important
difference in sources in the high-frequency and low-frequency bands:
whereas many (though certainly not all) high-frequency sources are
short-lived and comparatively rare (e.g., binary coalescence and
stellar collapse), most low-frequency sources are quite long-lived and
may not be so rare.  As in Sec.\ {\ref{subsec:high}}, we now take a
quick tour through some interesting LISA sources.

\subsubsection{Periodic emitters}

For LIGO, the source of most periodic gravitational waves is expected
to be isolated neutron stars, essentially gravitational-wave pulsars.
LISA's periodic sources will come primarily from binary star systems
in the Milky Way.  These systems do not generate waves strong enough
to backreact significantly on the system, so that their frequencies
typically change very little or not at all over the course of LISA
observations.  Certain systems are well-known in advance to be sources
of periodic waves for the LISA band; cf.\ the points in Fig.\
{\ref{fig:lisa_sens}}.  These sources are understood well enough that
they may be regarded as ``calibrators'' --- LISA had better detect
them, or else something is wrong!

Aside from these sources that are known in advance, it is expected
that LISA will discover a good number of binary systems that are too
faint to detect with telescopes.  Joint observations by LISA and other
astronomical instruments are likely to be quite fruitful, helping to
understand these systems much better than can be done with a single
instrument alone.  For example, it is typically difficult for
telescopes to determine the inclination of a binary to the line of
sight (a factor needed to help pin down the mass of the binary's
members).  Gravitational waves measure the inclination angle almost
automatically, since this angle determines the relative magnitude of
the polarizations $h_+$ and $h_\times$.

\subsubsection{Coalescing binary black holes}

Coalescing binary black hole systems will be measurable by LISA to
extremely large distances; even if such events are very rare, the
observed volume is enormous, so that an interesting measured rate
seems quite likely.  One class of such binaries consists of systems in
which the member holes are of roughly equal mass.  These binaries can
form following the merger of galaxies (or pregalactic structures)
containing a black hole in their core.  Depending on the mass of the
binary, the waves from these coalescences will be detectable to fairly
large redshifts ($z \sim 5 - 10$), possibly probing an early epoch in
the formation of the universe's structure.  (The optimal system mass
is near $10^5 - 10^6\,M_\odot$ --- the waves from smaller systems
aren't so loud, and so can't be measured quite as well; the waves from
larger systems come out at low frequencies where noise is strong.)
The rate at which such events are likely to occur, however, is
extremely uncertain.  It seems clear that, following the merger of
their host structures, the black holes will form a bound binary.  It
is not clear, however, whether this hole becomes bound tightly enough
that gravitational-wave emission importantly impacts its dynamics:
some simulations show that the binary ``stalls'' well before
gravitational waves become important {\cite{mm2001}}.  It is possible
that a later mechanism drives the holes closer together (see, for
example, Ref.\ {\cite{av2002}}); some observations hint that this in
fact may be happening {\cite{me2002}}.  If black hole mergers are
``efficient'' (there is roughly one binary black hole merger for every
merger of host structures), then the rate at which LISA measures these
events could be several per year {\cite{mhn2001}}.

The other major class of binary black hole systems consists of
relatively small bodies (black holes with mass $\sim 10\,M_\odot$,
neutron stars, or white dwarfs) that are captured by larger black
holes ($M\sim 10^5 - 10^7\,M_\odot$) such as are found at the cores of
many galaxies.  These extreme mass ratio binaries are created when the
smaller body is captured onto an extremely strong field, highly
relativistic orbit, generating strong gravitational waves.  Such
systems are measurable to a distance of a few gigaparsecs if the
inspiraling body is a $10\,M_\odot$ black hole, and to a distance of a
few hundred megaparsecs if the body is a neutron star or white dwarf.
LISA will measure the waves that come from the last year or so of the
smaller body's inspiral, probing the nature of the larger black hole's
gravitational field from deep within the hole's potential.  The rates
for such events are, again, not so well understood, depending in some
detail on the dynamical nature of the cores of galaxies.  Extremely
conservative estimates typically find that the rate of measurable
events for LISA should be at least several per year
{\cite{sigurdssonrees,sigurdsson}}.  Recent thinking suggests that
these rates are likely to be rather underestimated --- black holes
(which are measurable to much greater distances) are likely to
dominate the measured rate, perhaps increasing the rate to several
dozen or several hundred per year.

Both of these types of black hole binaries will be discussed in
greater depth in Sec.\ {\ref{sec:BBH}} and {\ref{sec:bothrodesy}}.

\subsubsection{Stochastic backgrounds}

As discussed in Sec.\ {\ref{subsec:high}}, ground-based detectors can
measure a stochastic background by correlating the data streams of
widely separated detectors.  LISA obviously cannot do this, since it
consists of a single antenna.  However, it can take advantage of a
different trick: by combining its six data streams in an appropriate
way, it can construct an observable that is completely {\it
in}sensitive to gravitational waves, measuring noise only
{\cite{aet99}}.  This makes it possible to distinguish between a
noise-like stochastic background and true instrumental noise, and
thereby to learn about the characteristics of the background
{\cite{hb2001}}.

The sensitivity of LISA will not be good enough to set interesting
limits on an inflationary gravitational-wave background: LISA will
only reach $\Omega_{\rm gw} \sim 10^{-11}$, about four orders of
magnitude too large to begin to say something about inflation
{\cite{turner1}}.  However, as was discussed in Sec.\
{\ref{subsec:high}}, LISA's band is well placed for other possible
sources of cosmological backgrounds.  In particular, waves generated
by the electroweak phase transition at temperature $T \sim 100 - 1000$
GeV would generate waves in LISA's band; they are likely to be
detectable if the phase transition is strongly first order (a scenario
that does not occur in the standard model, but is conceivable in
extensions to the standard model {\cite{kmk2001}}).  Likewise, LISA is
well-positioned to measure waves that may arise from extradimensional
dynamics in the early universe (depending rather strongly on the scale
of the extra dimensions {\cite{hogan_prl,hogan_prd}}).

\subsection{Ultra low frequency}
\label{subsec:ultralow}

The ultra low frequency band, $10^{-18}\,{\rm Hz}\lesssim f\lesssim
10^{-13}\,{\rm Hz}$, is better described by converting from frequency
to wavelength: for these waves, $10^{-5}\,H_0^{-1}\lesssim \lambda
\lesssim H_0^{-1}$, where $H_0^{-1}\sim 10^{10}$ light years is the
Hubble length today.  Waves in this band oscillate on scales
comparable to the size of the universe.  They are most likely to be
generated during inflation: quantum fluctuations in the spacetime
metric are parametrically amplified during inflation to relatively
high amplitude.  The rms amplitude to which the waves are amplified
depends upon the energy scale of inflation:
\begin{equation}
h_{\rm rms} \propto \left(E_{\rm infl}\over m_{\rm Planck}\right)^2\;.
\end{equation}
Measuring these inflationary gravitational waves would be a direct
probe of inflationary physics.  Detection of these waves has been
described as the ``smoking gun'' signature of inflation
{\cite{turner2}}.

During inflation, quantum fluctuations impact both the scalar field
which drives inflation itself (the inflaton $\phi$) and the metric of
spacetime.  These scalar and tensor perturbations, $\delta\phi(\vec
r,t)$ and $h_{ab}(\vec r,t)$, each satisfy a massless Klein-Gordon
equation.  The Fourier modes of each perturbation,
$\delta\tilde\phi(\vec k,t)$ and $\tilde h_{ab}(\vec k, t)$, are thus
describable as harmonic oscillators in the expanding Universe
{\cite{kamkos99}}.  Each mode undergoes zero-point oscillations in the
harmonic potential.  However, the potential itself is evolving due to
the expansion of the universe.  The evolution of this potential
parametrically amplifies these zero-point oscillations, creating
quanta of the field {\cite{bruce96}}.  During inflation, the scale
factor grows faster than the Hubble length $H^{-1}$, and so each
mode's wavelength likewise grows faster than the Hubble length.
Amplification of each mode occurs while its wavelength is smaller than
$H^{-1}$; when the scale factor has grown such that $\lambda\gtrsim
H^{-1}$, the crests and troughs of each mode are no longer in causal
contact and the fluctuation ceases to grow, becoming frozen at its
amplified magnitude {\cite{kamkos99}}.  Fluctuations in the inflaton
seed density fluctuations, $\delta\rho(\vec r) = \delta\phi(\vec
r)(\partial V/\partial\phi)$ [where $V(\phi)$ is the potential that
drives the inflaton field].  Fluctuations in the spacetime metric are
gravitational waves.

Both density fluctuations and gravitational waves imprint the cosmic
microwave background (CMB).  First, each contributes to the CMB
temperature anisotropy.  However, even a perfectly measured map of
temperature anisotropy cannot really determine the contribution of
gravitational waves very well because of {\it cosmic variance}: since
we only have one universe to use as our laboratory experiment, we are
sharply limited in the number of statistically independent influences
upon the CMB that we can measure.  Large angular scales are obviously
most strongly affected by this variance, and these scales are the ones
on which gravitational waves most importantly impact the CMB
{\cite{kamkos98}}.

Fortunately, the scalar and tensor contributions also impact the {\it
polarization} of the CMB.  These two contributions can be detangled
from one another in a model-independent fashion.  This detangling uses
the fact that the polarization tensor $P_{ab}({{\bf \hat n}})$ on the
celestial sphere can be decomposed into tensor harmonics.  These
harmonics come in two flavors, distinguished by their parity
properties: the ``gradient-type'' harmonics $Y_{(lm)ab}^G({\bf\hat
n})$ [which pick up a factor $(-1)^l$ under ${\bf\hat n}\to -{\bf\hat
n}$], and the ``curl-type'' harmonics $Y_{(lm)ab}^C({\bf\hat n})$
[which pick up a factor $(-1)^{l+1}$ under ${\bf\hat n}\to -{\bf\hat
n}$].  These harmonics are constructed by taking covariant derivatives
on the sphere of the ``ordinary'' spherical harmonics $Y_{lm}({\bf\hat
n})$; see {\cite{kks97}} for details.  (An alternative, but
equivalent, formulation labels the gradient-type harmonics ``E-modes''
and the curl-type harmonics ``B-modes'' {\cite{zs97}}; the analogy to
electric and magnetic fields is obvious.  Interestingly, the various
multipole formalisms used to describe polarization maps are identical
to those used to expand gravitational radiation fields, as in Ref.\
{\cite{thorne80}}; see Ref.\ {\cite{kks97}} for further discussion.)
Because scalar perturbations have no handedness, they {\it only}
induce gradient-type polarization.  Gravitational waves induce both
gradient- and curl-type polarization.  Thus, an unambiguous detection
of the curl-type polarization would confirm production of
gravitational waves by inflation.

The gradient-type polarization has recently been measured for the
first time {\cite{emodes}}.  These modes are reduced relative to the
CMB temperature anisotropy by an order of magnitude; the curl
component should be smaller by an additional order of magnitude
{\cite{hhz2002}}.  Detecting the gravitational-wave component of CMB
polarization will be quite a challenge --- aside from the instrumental
sensitivity needed to measure this effect {\cite{jkw2000}},
astrophysical foregrounds can cause important complications
{\cite{tetal2000,psb2000,baccietal2002}}, such as conversion of
gradient modes to curl modes {\cite{kck2002}}.  But this is likely to
be the only direct probe of physical processes in the inflationary
era.

\subsection{Very low frequency}
\label{subsec:verylow}

The very low frequency band, $10^{-9}\,{\rm Hz}\lesssim f\lesssim
10^{-7}\,{\rm Hz}$, corresponds to waves with periods ranging from a
few months to a few decades.  Our best limits on waves in this band
come from observations of millisecond pulsars.  First suggested by
Sazhin {\cite{sazhin}} and then carefully analyzed and formulated by
Detweiler {\cite{det79}}, gravitational waves can drive oscillations
in the arrival times of pulses from a distant pulsar.  The range
encompassed by the very low frequency band is set by the properties of
these radio pulsar measurements: the high end of the frequency band
comes from the need to integrate the radio pulsar data for at least
several months; the low end comes from the fact that we have only been
observing millisecond pulsars for a few decades.  One cannot observe a
periodicity shorter than the span of one's dataset!

Millisecond pulsars are very good ``detectors'' for measurements in
this band because they are exquisitely precise clocks.  Andrea Lommen
has recently {\cite{lommen2002}} performed a rather massive analysis
of the data from several millisecond pulsars that are widely spaced on
the sky.  Her analysis extends the data used for a previous analysis
{\cite{ktr94}} so that nearly 17 years of observations are
represented.  A detailed description of Lommen's methodology is given
in Ref.\ {\cite{lommen2002}}; her punchline is the following limit on
the density of stochastic gravitational waves:
\begin{equation}
\Omega_{\rm GW} h_{100}^2 < 2\times 10^{-9}
\end{equation}
(where $h_{100}$ is the Hubble constant in units of $100\,{\rm
km}\,{\rm sec}^{-1}\,{\rm megaparsec}^{-1}$).  This is the best
observed limit on gravitational waves that has been achieved to date.
Though it is not quite at the level where it can constrain sources of
stochastic gravitational-wave backgrounds, it is extremely close; with
further observations and the inclusion of additional pulsars in the
datasets, it is likely to become interesting quite soon.  It is
expected that the background in this band will be dominated by many
unresolved coalescing massive binary black holes {\cite{jb2002}} ---
binaries that are either too massive to radiate in the LISA band, or
else are inspiraling towards the LISA band en route to a final merger
several centuries or millenia hence.  Constraints from pulsar
observations in this band will remain an extremely important source of
data on stochastic waves in the future --- the limits they can set on
$\Omega_{\rm GW}$ are likely to be better than can be set by {\it any}
of the laser interferometric detectors.

\section{Binary black holes}
\label{sec:BBH}

As has been mentioned already in Secs.\ {\ref{subsec:high}} and
{\ref{subsec:low}}, one of the most important sources of gravitational
radiation in the high- and low-frequency bands is the coalescence of
compact objects.  One of the reasons for this importance is that this
source is amenable, at least to some degree, to fairly detailed
theoretical analysis: for the most part, the only tools needed to
understand the evolution of these systems are the nature of
gravitational-wave emission and the manner in which it drives these
binaries to coalesce.

Analysis of binaries becomes considerably more complicated when its
members come close together.  Then, the nature of these members can
become extremely important --- their finite size and the material of
which they are made importantly influences the binary's evolution and
the character of the waves that it generates.  For binaries that
contain neutron stars, the late stages of the ``inspiral'' (when the
members of the binary are well separated and evolve primary due to
gravitational-wave backreaction) and the final ``merger'' (when the
bodies come into contact and fuse into some kind of remnant) will
depend in detail on the nature of neutron star matter
{\cite{zcm,rs,fgrt,shibata,wigginslai}}.

The problem remains ``clean'', at least in principle, if both members
of the binary are black holes.  There is then no matter to complicate
the problem --- black holes are vacuum solutions to the Einstein field
equations, and so a binary black hole system is likewise just a vacuum
solution.  The dynamics of binary black holes can be stated quite
concisely: they are given by the family of dynamical spacetimes,
$g_{ab}(t)$, which: (a) satisfy the vacuum Einstein field equations
$G_{ab} = 0$; (b) consist of a pair of widely separated black holes in
the asymptotic past; (c) consist of a single rotating black hole in
the asymptotic future; (d) allow only outgoing radiation to reach
distant observers (who are located at ``outgoing null infinity''); and
(e) allow only ingoing radiation to propagate down event horizons.
[For careful definitions of the Einstein tensor $G_{ab}$ and outgoing
null infinity, see, e.g.\ {\cite{wald}}.  Note that the time parameter
$t$ introduced in the metric $g_{ab}(t)$ is intended to be any
future-directed label that parameterizes the evolution of the system.
For the purposes of gravitational-wave astronomy, a convenient such
label is time measured by very distant observers --- i.e., us.]

As is often the case in mathematics, the ease with which the problem
can be stated belies the difficulty one has in solving it.  The field
equation $G_{ab} = 0$ is shorthand for ten coupled nonlinear partial
differential equations.  The location of event horizons (upon which
one might naively want to place the ``ingoing radiation only''
boundary condition) is not known in advance, and as a matter of
principle {\it cannot} be known until the full spacetime is built.
And, in general relativity one has a great deal of freedom to specify
coordinates.  It is not often clear, for the purposes of a
calculation, what particular choice will turn out to be ``good''.
Despite their ``clean'' character, binary black hole systems are not
at all easy to describe.

A useful (albeit very crude) characterization of binary systems breaks
their evolution into three broad epochs.  The characterization that we
will use here in based on that presented in Ref.\ {\cite{fh1998}}; as
we will discuss further below, there is a fair amount of arbitrariness
associated with this characterization.  The first two epochs have
already been mentioned: the {\it inspiral} describes the binary when
its members are separated, discrete objects, evolving primarily due to
the backreaction of gravitational-wave emission.  The {\it merger}
which follows describes the violent dynamics of the two bodies merging
into a single body.  For binary black hole systems, this remnant will
itself be a black hole.  (The remnant most likely will contain a black
hole for binaries with neutron stars as well.)  This remnant hole must
``settle down'' to the Kerr solution {\cite{kerr}} which describes all
rotating black holes --- the ``no hair'' theorem of general relativity
{\cite{nohair}} guarantees that the Kerr solution describes the final
state, no matter what conditions describe the binary which produced
it.  This ``settling down'' process has been named the {\it ringdown}
since the waves generated in this epoch take the form of damped
sinusoids, similar to the sound of a struck bell.  In fact, the
quality factor $Q$ of black holes is quite low ($Q_{\rm BH} \sim 20$
or so, compared to $Q_{\rm bell} \sim 10^3-10^5$); when translated
into sound, one finds that black holes don't ring so much as thud
{\cite{thud}}.  Ringdown waves ``shave'' the remnant, ensuring that
all of the ``hairiness'' characterizing the system right after the
merger is lost, and what remains is a perfectly hairless Kerr black
hole {\cite{price1,price2}}.

Breaking the coalescence process into three broad epochs likewise
divides its gravitational waves into three broad frequency bands.
(This is one reason that this characterization is useful, despite its
crudeness --- it illustrates what source dynamics are ``audible'' to
the observatories.)  Roughly speaking, for inspiral waves we have
{\cite{fh1998}}
\begin{equation}
f \lesssim 400\,{\rm Hz}\left[{10\,M_\odot\over(1 + z)M}\right]\;,
\label{eq:f_inspiral}
\end{equation}
where $z$ is the cosmological redshift and $M$ is the total system
mass.  The ringdown waves come out at frequency
\begin{eqnarray}
f &\sim& {c^3\over2\pi G(1 + z)M}\left[1 + 0.63(1 - a/M)^{0.3}\right]
\nonumber\\
&\sim& \left(1200 - 3200\right)\,{\rm Hz}\left[{10\,M_\odot\over(1 +
z)M}\right]\;.
\label{eq:f_ringdown}
\end{eqnarray}
The parameter $a$ describes the spin of the merged remnant: it is
related to the vectorial black hole spin ${\vec S}$ by $a \equiv
G|{\vec S}|/Mc$, and is in the range $0 \le a \le M$.  The span in
frequency given in Eq.\ (\ref{eq:f_ringdown}) reflects this range.
These ringdown waves are generated by a bar-like perturbation to the
black hole that rotates in the same sense as the hole's spin.  The
``merger'' then consists of all waves that come out between these two
frequencies.

This division into three bands, particularly our definition of the
``merger'', is rather crude and ad hoc.  The notion of ``inspiral'' is
wholly defensible when the holes which comprise our binary are widely
separated.  The binary's dynamics are then well described using the
post-Newtonian approximation to general relativity
{\cite{blanchet_lr}}: the lowest order dynamics are described by
Newtonian gravity, and corrections to this motion are given in terms
of a power series in $x \sim (G M/rc^2)^{1/2}$, where $r$ is orbital
separation.  The parameter $x$ is roughly orbital speed over $c$.
This expansion works well when $x$ is small.  Late in the inspiral,
when $x \sim 0.2 - 0.4$, the convergence of this power series is not
so good.  The frequency given in Eq.\ (\ref{eq:f_inspiral})
corresponds roughly to this $x$. (Further discussion and caveats can
be found in Sec.\ III of Ref.\ {\cite{fh1998}}.)  Likewise, the notion
of ``ringdown'' is quite rigorous and defensible as a means of
describing the last waves that flutter out of the merged system ---
the remnant of the binary can be treated as a Kerr black hole plus
some distortion; perturbation theory accurately describes the waves
generated in this state {\cite{teuk73,leaver}}.  This is in fact how
Eq.\ (\ref{eq:f_ringdown}) was found {\cite{leaver,echeverria}}.

Difficulties come in the middle: what we have called ``merger'' sweeps
together all of the poorly understood physics associated with the end
of the inspiral and the complex gravitational dynamics describing the
transition of our binary into a single black hole.  Note that, for
binaries of several tens of solar masses, the frequencies associated
with these poorly understood waves lie very near the most sensitive
frequencies of ground-based gravitational wave detectors.  These
waves, which we currently understand least well, may be perfectly
suited for gravitational-wave observatories to measure!

This is the vanguard of current research in binary systems in general
relativity, motivated quite a bit by the likely observational
importance of the late inspiral and merger waves.  Much of the
community's efforts to understand strong-field binary black hole
dynamics use {\it numerical relativity}: direct solution of the
Einstein field equations by large scale computations.  In principle,
numerical relativity should be able to provide, in detail, a
description of the binary's dynamics as a function of the two black
holes' masses and spins, and thus the gravitational waveforms produced
by these dynamics.  These waveforms should depend uniquely on these
masses and spins since they are the only parameters that can describe
the binary's holes.  Comparison of the numerically generated waveform
with those measured by gravitational-wave observatories is arguably
the most stringent test of general relativity imaginable, probing what
are probably the strongest and most violently varying gravitational
fields produced by nature since the big bang.

Numerical solution of the two black hole problem has proven to be
quite difficult.  Unanticipated problems have slowed the rate of
progress in this field to the point that astrophysically relevant
binary solutions are just beginning to be produced today.  Some idea
of how unanticipated these problems were can be inferred from the
following statement by Kip Thorne:
\begin{quote}
...numerical relativity is likely to give us, in the next five years
or so, a detailed and highly reliable picture of the final coalescence
and the wave forms it produces, including the dependence on the holes'
masses and angular momenta.
\end{quote}
This statement was written in a well-known review article from 1987
(Ref.\ {\cite{300yrs}}, p.\ 379); clearly, Thorne's estimate of the
timescale needed to get out interesting information was optimistic.

Many of the most important problems are beginning to be understood ---
progress in numerical relativity has been quite impressive recently.
We will just summarize some of the recent highlights; the interested
reader will find more details in the review by Lehner, Ref.\
{\cite{lehner}}.  One of the fundamental difficulties has been casting
Einstein's equations into a form that behaves well under numerical
integration.  Some formulations which behaved quite well on earlier
testbed problems with high degrees of symmetry have been found to
perform extremely badly in general {\cite{kst2001}}: they allow
unphysical modes (which are seeded by very small scale numerical
errors) to grow exponentially and destroy the physical content of a
calculation.  Understanding this behavior will hopefully make
controlling it possible, so that we will be able to construct
evolution schemes that are not susceptible to unphysical mode growth
{\cite{ls2002,sklpt}}.

Despite the fact that codes currently cannot model the full binary
black hole merger right now, success has been achieved by taking
present codes as far as they can go and then using perturbation theory
to carry the evolution still further.  This very pragmatic approach
takes the point of view that the ``full'' codes should only be used
for a limited section of the merger process {\cite{bcc}}.  Dubbed
``The Lazarus Project'' (since it works by resurrecting a fallen
code), this direction makes it possible to get some insight into the
properties of the waves generated late in the merger process
{\cite{bclt}}.

Even with good evolution equations and perfect codes, it is necessary
to match the strong-field portion of the coalescence which has been
numerically modeled to the earlier inspiral --- the initial data with
which one starts the numerical evolution must latch onto what came
before.  It now seems likely that such data will be well-developed
fairly soon.  A way to approximate an evolution is to consider it to
be a sequence of initial data snapshots.  This works well provided
that the evolution of the system is not too rapid --- the binary can
be treated as in {\it quasi-equilibrium}.  Such techniques were
originally developed to study binary neutron star systems
{\cite{wmm,bcsst,bgm,shibata98,ue2000}}.  Recently an extension to
this technique has been developed which goes beyond the ``slices of
initial data'' view, endowing the spacetime with a helical timelike
Killing vector which describes with good accuracy the circular motion
of binary black holes {\cite{ggb1,ggb2}}.  With these tools, it should
not be too difficult to go from the earlier inspiral regime into the
very strong field merger, covering the full range of binary black hole
coalescence.

In parallel to the recent progress in numerical relativity, techniques
have been developed by Thibault Damour and colleagues
{\cite{bd99,bd2000,djs2000,d2001}} that promise to greatly improve our
{\it analytical} understanding of strong-field binary systems.  This
work is based on combining ``resummation methods'' to improve the
post-Newtonian description of the binary with a novel recasting of the
binary's dynamics in terms of the motion of a single body in an
``effective one-body metric'' (usefully regarded as a deformed black
hole).  The resummation techniques are, essentially, Pad\'e
approximants that improve the behavior of the poorly convergent Taylor
series form of the post-Newtonian expansion.  The one-body remapping
is based on tools that were originally developed to describe two-body
problems in quantum electrodynamics; further discussion can be found
in Ref.\ {\cite{bd99}}.  Good agreement has been found between
important invariant dynamical quantities describing strong-field
binary orbits using this effective one-body technique and numerical
relativity {\cite{dgg}}.

These rapidly maturing approaches to strong field dynamics gives us
hope that theory will be able to play an important role aiding and
interpreting gravitational-wave observations of black hole binaries.
As has already been mentioned above, comparing measured binary black
hole waves to those predicted by theory is about the most stringent
test of general relativity imaginable.  In addition to this ``physics
measurement'', the waves will provide a wealth of astrophysical
information.  As discussed in Secs.\ {\ref{subsec:high}} and
{\ref{subsec:low}}, we currently know very little about the rate at
which these mergers are likely to take place.  Any information about
the rate will provide a great deal of information: observations in the
high-frequency band by LIGO-type instruments can strongly constrain
the various scenarios (e.g., Refs.\
{\cite{bb98,pzy98,fwh99,pzw2000,bkb2002,nstt97,ictn98}}) by which
stellar mass binaries can form; observations with LISA may be able to
directly observe the consequences of early hierarchical mergers that
were the building blocks of galaxies {\cite{mhn2001}}.

Detailed information about the binary that generates a particular
signal will be measurable in cases in which we can fit the data to a
model waveform --- such fits provide (with varying degrees of
accuracy) certain combinations of the black holes' masses, information
about their spins, the source's position on the sky, and the distance
to the source (cf.\ Refs.\
{\cite{cutler98,echeverria,fc93,cf94,pw95,untangle}} for further
discussion).  This information greatly increases the astrophysical
value of gravitational-wave measurements.  For example, using LISA it
should be possible to survey the evolution of black hole masses as a
function of redshift {\cite{untangle}}, tracing the development of
black holes and the structures that host them over the evolution of
the universe.  If an electromagnetic counterpart can be associated
with the gravitational-wave event, the measurement could provide a
standard candle with extraordinarily low intrinsic error
{\cite{BBHcandle}}.

Though much is unknown about binary black holes in the universe, it is
clear they are exquisite gravitational-wave sources --- they are
intrinsically ``loud'' radiators, they are incredible labs for testing
gravity under extreme conditions, and they are powerful probes of
astrophysical processes.

\section{Bothrodesy}
\label{sec:bothrodesy}

One subset of binary black holes comprises a LISA source with
particularly wonderful characteristics.  These are the extreme mass
ratio binaries mentioned in Sec.\ {\ref{subsec:low}} --- binaries
formed by the capture of stellar mass compact objects onto highly
relativistic orbits of massive black holes.  (As described in Sec.\
{\ref{subsec:low}}, the captured object can be a neutron star or a
white dwarf as well as a black hole.  Since black holes are likely to
dominate the measured rate, we will consider this source to be a
special case of binary black holes.)

In the general case, the spacetime of a binary black hole is a
violently dynamical entity, varying in a manner that is extremely
difficult to model (cf.\ the discussion in Sec.\ {\ref{sec:BBH}}).
The character of extreme mass ratio binaries is quite different.
Because the captured body is so much less massive than the large black
hole, the binary's spacetime is largely that of the black hole plus a
perturbation.  The major effect of this perturbation is to create
gravitational radiation.  The motion of the small body is essentially
an orbit that evolves due to this radiation.  The properties of this
evolving orbit --- and thus of the waves that it generates --- depend
almost entirely on just the large black hole's spacetime.  These waves
provide an extremely clean probe of the black hole's spacetime.

Einstein's theory of gravitation predicts that black holes are objects
with event horizons, and whose structure is completely described by
two numbers, the mass $M$ and spin parameter $a$ (ignoring the
astrophysically uninteresting possibility of a charged black hole ---
macroscopic charged objects are rapidly neutralized in astrophysical
environments by interstellar plasma).  Extreme mass ratio inspirals
provide a way to test this: the gravitational waves generated as the
compact body spirals through the strong field of the black hole depend
upon, and thus encode, the structure of the hole's spacetime metric.

The waves that LISA will measure come from the captured body spiraling
through the very strong field of the large black hole --- the orbital
radius is a few times the Schwarzschild radius of the hole, so that
the captured body is near the hole's event horizon.  The small body
executes many orbits as gravitational-wave backreaction drives it to
spiral inwards --- it orbits about $10^5 - 10^6$ times before it
reaches a dynamical instability and then plunges into the hole.  These
orbits happen over a period of several months to years.  By tracking
the gravitational wave's phase evolution over this time, we will be
able to follow the evolution of the smaller body's orbital frequencies
with high precision.

It is these frequencies, or rather the sequence of frequencies that
the small body follows, which encode such information about the black
hole spacetime.  Consider for a moment an eccentric, inclined orbit
about a spherical body with mass $M$.  The concept of ``inclination''
is of course rather artificial in this case --- the field will be
spherically symmetric, so the orbits had better not depend on that
inclination.  Ignoring this common sense for a moment, we can define
three orbital timescales: $T_r$ is the time it takes to move through
the full range of motion in the radial coordinate; $T_\theta$ is the
time it takes to move through the full range of latitudinal angle; and
$T_\phi$ is the time it takes to move through $2\pi$ radians of
azimuth.

For spheres in Newtonian gravity, these three timescales are of course
identical: $T_r = T_\theta = T_\phi \equiv T = 2\pi\sqrt{R^3/M}$ ---
Newtonian orbits are closed ellipses with semi-major axis $R$.  That
$T_\theta = T_\phi$ follows from the spherical symmetry of the
gravitational field.  That $T_r$ is equal as well is something of a
miracle that follows from the $1/r$ form of Newton's gravitational
potential {\cite{bertrand}}.  Now imagine adding some multipolar
structure to the sphere.  This changes the character of the potential,
and thus the character of the frequencies.  For example, if we add a
quadrupolar distortion to our sphere, the gravitational potential
picks up a bit that goes as $1/r^3$ and that has an angular
dependence:
\begin{equation}
V_{\rm grav} = - {G M \over r} + {{\mathcal Q}Y_{20}\over r^3}\;.
\label{eq:quad_newt}
\end{equation}
($\mathcal Q$ heuristically represents the quadrupolar distortion of
the central body; $Y_{20}$ is a spherical harmonic.)  This extra piece
changes all of the timescales --- we no longer have $T_\phi =
T_\theta$ for example, because the potential is no longer spherical.

Measuring the orbital frequencies thus maps the shape of a body's
gravitational field, which in turn maps the body's structure.  Using
satellite orbits, we have measured with high precision quite a few of
the multipolar distortions that characterize the Earth; NASA's
recently launched GRACE mission {\cite{grace}} promises to improve
these measurements quite a bit (see {\cite{phystoday}} for further
discussion).  The science of performing these measurements is known as
{\it geodesy}.

In a very similar way, by tracking the evolution of the orbital
frequencies that describe black hole orbits through the gravitational
waves that they generate, we can map the shape of a black hole's
spacetime metric.  In analogy to geodesy, this science has been given
the name {\it bothrodesy}.  This name comes from the Greek word
``bothros'' (${\beta}o{\theta}\!{\rho}o{\varsigma}$), meaning
(roughly) ``garbage pit''.  (In archeology, ``bothros'' refers to a
sacrificial pit --- an appropriate connotation since a black hole is
Nature's ultimate sacrificial pit!)

Bothrodesy is particularly powerful because black holes have a unique
multipolar structure.  As we have already stated, the ``no-hair''
theorem {\cite{nohair}} tells us that the spacetime of a black hole
can only depend on its mass $M$ and spin $a$.  On the other hand, it
is well understood that the spacetime of a compact object can be built
from a multipolar description of that object {\cite{thorne80}}.  The
object is fully described by a family of mass moments $M_{lm}$
(similar to electric multipole moments) and current moments $S_{lm}$
(analogous to magnetic multipole moments) given roughly by
\begin{eqnarray}
M_{lm} &\simeq& \int dV\,r^l\,Y_{lm}(\theta,\phi)\,
\rho(r,\theta,\phi)\;,
\\
S_{lm} &\simeq& \int dV\,r^l\,Y_{lm}(\theta,\phi)\,
\rho(r,\theta,\phi)v(r,\theta,\phi)\;;
\end{eqnarray}
$\rho$ is the mass density at the coordinate $(r,\theta,\phi)$, and
$\rho v$ is the current density.  Although a black hole has no matter,
its spacetime is also generated by multipole moments of this form.
The moments of a black hole are
\begin{eqnarray}
M_{l0} &+& i S_{l0} = M(ia)^l\;,
\label{eq:moments}
\\
M_{lm} &=& S_{lm} = 0\qquad\mbox{for $m\ne0$}\;.
\label{eq:zero_moments}
\end{eqnarray}
Condition (\ref{eq:zero_moments}) simply enforces the fact that
rotating black holes are axisymmetric.  Condition (\ref{eq:moments})
is far more interesting: it enforces the no-hair theorem!  For $l =
0$, it tells us $M_{00} = M$ --- the zeroth mass moment is the mass,
no great surprise.  For $l = 1$ we find $S_{10} = a M$.  This is the
magnitude of the hole's spin $|\vec S| = S$ (in units with $G = 1 =
c$).  {\it All higher multipoles are completely determined by these
first two moments.}

This is a remarkably powerful statement.  It tells us that measuring
three multipole moments is sufficient to falsify whether an object is
a black hole.  For example, many galaxies are known to contain
extremely massive, compact gravitating objects in their centers.  It
is most plausible that these objects are black holes, but it is
possible they could be something even more bizarre, such as a
gravitational condensation of bosonic cold dark matter
{\cite{csw86,lm92,jetzer92,lp92,ds2000}}.  If we measure gravitational
waves from inspiral into one of these massive objects and find that
the moment $M_{20}$ is not consistent with the measured values of $M$
and $S$, then that object is {\it not} in fact a black hole, but is
indeed something even more bizarre.  Conversely, if we can measure a
good sized set of multipoles and find that they are all consistent
with Eq.\ (\ref{eq:moments}) then we have extremely compelling
evidence that the ``black hole'' is in fact a black hole exactly as
described by general relativity.

How well can LISA perform this kind of measurement in practice?
Laying the foundations to answer this question is an area of very
active research right now.  Some guidance can be found from
calculations performed by Fintan Ryan {\cite{fintan}}.  Ryan examined
how well one can measure the moment structure of a large body with
gravitational waves in the context of a toy calculation.  In his
setup, the inspiraling body is confined to orbits that lie in the
large body's equatorial plane and are of zero eccentricity.  These
restricted orbits throw away a lot of useful information about the
multipolar structure which would be encoded in the precessional motion
of an orbit that is inclined and eccentric.  Ryan's calculation
instead ``weighs'' the different multipoles by the fact that each
impacts the orbital frequency with a different radial dependence, and
so affects the waveform phasing at different rates as the small body
spirals in.  Even in the context of this excessively simplified
problem, Ryan finds that at least three and in some cases five
multipoles will be measurable by LISA.  We are certain that, due to
his restricted orbit families, Ryan's calculation underestimates how
well LISA will be able to measure these moments.

It's worth noting at this point the accuracy with which some of these
moments can be measured.  Ryan finds {\cite{fintan}} that the mass of
the large object is typically measured with an accuracy $\delta M/M
\sim 10^{-4} - 10^{-5}$.  This is phenomenal precision --- the
precision with which we measure black hole masses today is no better
than $\sim10\%$ for the Milky Way's black hole, and usually much
larger ($\delta M/M \sim 1$ or larger is not uncommon).  Ryan finds
that the spin can be measured with an accuracy $\delta S/S \sim 0.01$.
This again is extremely precise --- presently, we have very little
information about black hole spins, other than indications that the
spin must be rapid in some cases {\cite{wilms,elvis}}.  It's worth
re-emphasizing that his accuracy estimates are likely to be
pessimistic owing to his excessively restricted orbit families.
Bothrodesy will provide high precision probes of the nature of black
holes.

Preparing for these LISA observations requires that we understand the
nature of the waves that inspiral into black holes will provide.
Because of the extreme mass ratio of inspiral systems, this is a
relatively simple task: black hole perturbation theory using the
system's mass ratio as an expansion parameter describes these binaries
very well
{\cite{poisson93,cfps,akop,ckp,dank98,paperI,paperII,zoomwhirl}}.
Although there remain issues of principle that are currently being
worked out (particularly the issue of rigorously computing the
perturbation's backreaction on the inspiral in full generality
{\cite{mst97,qw97,wiseman2000,bo2000,burko2000,lousto2000,bmnos,pp2002}}),
this problem is not nearly as difficult as that of the general binary
black hole evolution.  Indeed, there are two special cases in which
perturbative codes have already been able to tell us a great deal
about the character of these inspirals.  These cases correspond to
orbits that are ``circular'' but inclined, and orbits that are
eccentric but confined to the hole's equatorial plane.

\begin{figure}[t]
\includegraphics[width = 13cm]{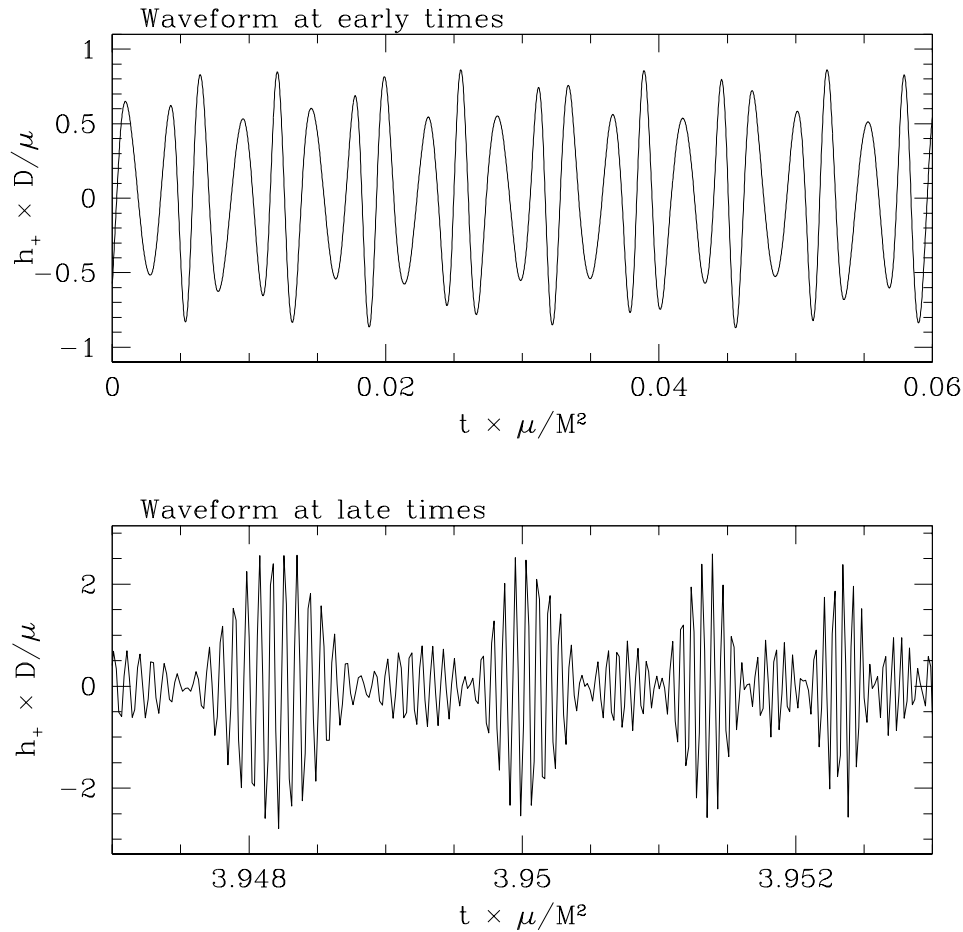}
\caption{The waveform generated by ``circular'' inspiral, from Ref.\
{\cite{paperII}}.  Early on, the modulation is small and happens on a
short timescale.  This is because the frequencies $\Omega_\phi$ and
$\Omega_\theta$ describing circular motion are not very different.
The frequencies evolve at different rates, changing the nature of the
modulation dramatically as time proceeds.  At late times, the
modulation is very strong, and there are many more cycles of
``carrier'' in each cycle of modulation.  Note the different
timescales in the top and bottom panels --- orbital frequencies are
much higher late in inspiral.  Audio encodings of this waveform can be
downloaded from {\cite{sounds}}.
}
\label{fig:circ_insp}
\end{figure}

Let us look at the circular inspirals first.  Circular orbits would be
of constant radius if radiative backreaction were not shrinking them.
Waveforms generated in this case are influenced by two orbital
frequencies, $\Omega_\phi$ (related to the time required for an orbit
to move through $2\pi$ radians of azimuth) and $\Omega_\theta$
(related to the time required to span its full range of latitude).
These frequencies differ for rotating black holes, in part because
rotation makes black holes oblate [cf.\ the discussion near Eq.\
(\ref{eq:quad_newt})] and in part because of {\it frame dragging} ---
the tendency of objects near a spinning source of gravity to be
dragged into corotation with that spin.  Under the combined influence
of these two effects, $\Omega_\phi > \Omega_\theta$.  The difference
leads to a modulation of the gravitational waveform --- essentially,
there is beating between these two frequencies.

This modulation is illustrated in Fig.\ {\ref{fig:circ_insp}} (taken
from Ref.\ {\cite{paperII}}).  Here we show an example of an inclined,
circular inspiral into a rapidly rotating black hole (spin parameter
$a = 0.998 M$).  Segments of the waveform are presented early in the
inspiral and again much later (as the inspiraling body approaches the
final plunge orbit).  Note the evolving character of the waveform's
modulation: the amplitude of the modulation is much stronger at the
end, and there are many more cycles of the carrier wave per cycle of
the modulation.  This is a signature of the black hole's strong field:
near the event horizon, $\Omega_\theta$ decreases (a redshifting
effect due to the proximity of the event horizon), whereas
$\Omega_\phi$ grows to a maximum (the body ``locks'' onto the dragging
of inertial frames and is forced to orbit at a rapid rate
{\cite{membrane}}).  In the physical space near the hole, the small
body appears to whirl very rapidly near the black hole while slowly
moving in its latitude angle.  This stamp on the waveform is a clear
signature of a black hole's strong field nature.

Eccentricity introduces yet another layer of complexity, owing to
modulations between the inspiraling object's azimuthal motion and its
motion in the radial direction.  Strong-field eccentric orbits show
what has been named a ``zoom-whirl'' character {\cite{zw_curt}}.  If
gravity were purely Newtonian, the inspiraling body would accumulate
$2\pi$ radians of azimuth while moving through its full range of
radius.  General relativity tells us that in fact the body moves
through an extra bit of azimuth over the orbit.  This effect is
nothing more than perihelion precession, well-known from studies of
Mercury's orbit in the solar system.

\begin{figure}[t]
\includegraphics[width = 13cm]{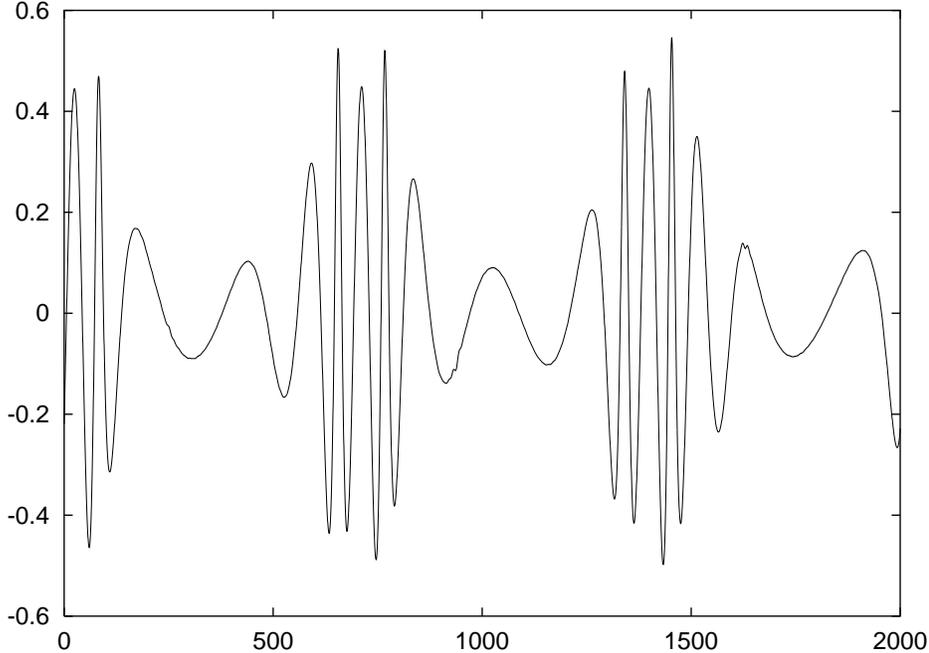}
\caption{A ``zoom-whirl'' waveform, generated by an eccentric,
equatorial orbit, from Ref.\ {\cite{zoomwhirl}}.  The high frequency
peaks near $t\sim0$, $t\sim700$, and $t\sim1400$ are due to the
whirling motion of the inspiraling body at peribothron.  This is a
relatively gentle zoom-whirl structure --- it is not difficult to find
cases that exhibit stronger whirling at peribothron.  Audio encodings
of waveforms that incorporate this kind of structure can be heard at
{\cite{sounds}}.
}
\label{fig:zoomwhirl}
\end{figure}

In the case of Mercury, the excess azimuth is rather puny --- an extra
$43$ arcseconds of azimuth accumulate every century due to general
relativity, or about $0.1$ arcsecond per orbit.  In the strong field
of a rapidly rotating black hole, the extra azimuth can amount to
thousands of degrees per orbit!  The inspiraling body appears to
``whirl'' around the black hole many times when it is near peribothron;
it then ``zooms'' out to apobothron and back, to whirl again on the next
cycle.  An example of the waveform from such an orbit (taken from
Ref.\ {\cite{zoomwhirl}}) is shown in Figure {\ref{fig:zoomwhirl}}.
Note the multiple high frequency cycles occurring every $t\sim 700$;
this is due to the rapid whirling of the inspiraling body at
peribothron.

The ornate character of the waves illustrated in Figs.\
{\ref{fig:circ_insp}} and {\ref{fig:zoomwhirl}} gives some sense of
the information that they encode.  These figures don't really do the
waveform justice, though --- to really get a sense of their harmonic
content, one should listen to an audio encoding of these waves.  The
reader is invited to listen to such encodings which have been placed
on the World Wide Web at the URL given in {\cite{sounds}}.  The sounds
presented there illustrate a variety of extreme mass ratio inspiral
signals, and how their features vary as a function of the system's
parameters.

\section{Conclusions}

In this article, we have taken a brief tour of various ways that the
Universe produces gravitational waves, surveying the different bands
in which this ``voice'' operates, and how we can build --- or are
building --- ``ears'' for listening to what it is saying.  Sections
{\ref{sec:BBH}} and {\ref{sec:bothrodesy}} have focused on the waves
produced from black hole sources, a particular favorite of this
author, outlining the challenges in learning to speak the language of
these sources and showing a few snippets of what we have learned so
far.

Before too long, we will hopefully begin to hear these voices directly
from Nature, and not just as output from theorists' computations.

\section*{Acknowledgments}

In presenting an overview of gravitational-wave astrophysics, I have
tried to emulate the style by which I learned the subject from Kip
Thorne.  I also thank Kip for his hospitality, without which I
probably would never have finished this article, as well as for
permission to reproduce Figs.\ {\ref{fig:forcelines}} and
{\ref{fig:ifo}} here.  I thank Kostas Glampedakis and Daniel Kennefick
for permission to use Fig.\ {\ref{fig:zoomwhirl}} (taken from Ref.\
{\cite{zoomwhirl}}).  Finally, I thank Daniel Holz for providing
useful comments on a careful, critical reading of this paper.  Some of
the background material presented here was adapted from a previous
review, Ref.\ {\cite{snowmass}}.  This work was supported by NSF Grant
PHY--9907949.


\begin{thebibliography}{99}

\bibitem{barish_ucsb} This very useful analogy between gravitational
  waves and the neutrino was described in a colloquium given by Barry
  Barish (Director of LIGO) at UC Santa Barbara on 7 May 2002.

\bibitem{chiao} Raymond Chiao has proposed that certain
  superconducting states may strongly couple to gravitational fields
  and thus may work as both antennae and generators of very high
  frequency gravitational waves; see gr-qc/0208024 for a review.
  Though the gravitational-wave detection community regards Chiao's
  ideas somewhat skeptically, as of the writing of this review they
  have not been solidly rebutted.  These ideas are certainly an
  interesting contender for a ``Savannah River'' type of
  gravitational-wave source.

\bibitem{radioastronomy_note} A counterexample to this is found in
  certain radio astronomy measurements, which can measure a coherent
  electromagnetic radiation field, just as gravitational-wave detectors
  measure a coherent gravitational radiation field.  (I thank Neil
  Cornish for pointing this out to me.)

\bibitem{sounds} {\tt
  http://www.tapir.caltech.edu/\~\/hughes/Inspiral/inspiral.html}

\bibitem{marcia} M.\ Bartusiak, ``Einstein's Unfinished Symphony:
  Listening to the Sounds of Space-Time'', Joseph Henry Press,
  Princeton, 2000.

\bibitem{laplace} P.\ Laplace, ``Sur le Principe de la Gravitation
  Universelle'', 1776; reprinted in ``Ouevres compl\'etes de Laplace
  VIII'', p.\ 201, Gauthier-Villars et fils, Paris, 1891.

\bibitem{dank_thesis} D.\ Kennefick, ``Controversies in the History of
  the Radiation Reaction Problem in General Relativity'', unpublished
  Ph.\ D.\ thesis (Part II), California Institute of Technology, 1997.

\bibitem{poincare} H.\ Poincar\'e, ``La dynamique de l'electron'',
  Revue g\'en\'erale des sciences pures et appliqu\'es {\bf 19}
  (1908), 186; reprinted in ``Ouevres de Henri Poincar\'e IX'', p.\
  551; translated by F.\ Maitland as ``The New Mechanics'', in
  ``Science and Method'', p.\ 199, Dover, New York, 1952.

\bibitem{bigal0} A.\ Einstein, K\"oniglich Preu{\ss}ische Akademie der
  Wissenschaften Berlin, Sitzungsberichte (1915), 831.

\bibitem{bigal1} A.\ Einstein, K\"oniglich Preu{\ss}ische Akademie der
  Wissenschaften Berlin, Sitzungsberichte (1916), 688.

\bibitem{bigal2} A.\ Einstein, K\"oniglich Preu{\ss}ische Akademie der
  Wissenschaften Berlin, Sitzungsberichte (1918), 154.


\bibitem{damour_prl} T.\ Damour, Phys.\ Rev.\ Lett.\ {\bf 51} (1983),
  1019.

\bibitem{hulse_taylor} R.\ A.\ Hulse and J.\ H.\ Taylor, Astrophys.\
  J.\ {\bf 195} (1975), L51.

\bibitem{taylor_weisberg} J.\ H.\ Taylor and J.\ M.\ Weisberg,
  Astrophys.\ J.\ {\bf 345} (1989), 434.

\bibitem{blanchet_lr} L.\ Blanchet, ``Gravitational Radiation from
  Post-Newtonian Sources and Inspiralling Compact Binaries'', Living
  Rev.\ Relativity {\bf 5} (2002), 3.\\ Online Article, cited on 14
  October 2002:\\ {\tt
  http://www.livingreviews.org/Articles/Volume5/2002-3blanchet/}

\bibitem{snowmass} S.\ A.\ Hughes, S.\ Mark\'a, P.\ L.\ Bender, and
  C.\ J.\ Hogan, {\it in} ``Proceedings of the APS / DPF / DPB Summer
  Study on the Future of Particle Physics (Snowmass 2001)'', (R.\
  Davidson and C.\ Quigg, Eds.), eConf C010630 (2001) P402; also
  astro-ph/0110349.

\bibitem{ligoscience} A.\ Abramovici et al., Science {\bf 256} (1992),
  325.

\bibitem{300yrs} K.\ S.\ Thorne, {\it in} ``300 Years of Gravitation''
  (S.\ W.\ Hawking and W.\ Israel, Eds.), p.\ 330, Cambridge
  University Press, Cambridge, 1987.

\bibitem{nps} R.\ Narayan, T.\ Piran, and A.\ Shemi, Astrophys.\ J.\
  {\bf 379} (1991), L17.

\bibitem{phinney91} E.\ S.\ Phinney, Astrophys.\ J.\ {\bf 380} (1991),
  L17.

\bibitem{kl2000} V.\ Kalogera and D.\ R.\ Lorimer, Astrophys.\ J.\
  {\bf 530} (2000), 890.

\bibitem{weiss72} R.\ Weiss, Quarterly Progress Report of RLE, MIT
  {\bf 105} (1972), 54.

\bibitem{pirani} F.\ A.\ E.\ Pirani, Acta Physica Polonica {\bf 15}
  (1956), 389.

\bibitem{gp1962} M.\ E.\ Gertsenshtein and V.\ I.\ Pustovoit, Soviet
  Physics --- JETP {\bf 16} (1962), 433.

\bibitem{drever} R.\ W.\ P.\ Drever, {\it in} ``Gravitational
  Radiation'' (N.\ Deruelle and T.\ Piran, Eds.), North Holland,
  Amsterdam, 1983.

\bibitem{levin} Yu.\ Levin, Phys.\ Rev.\ D {\bf 57} (1998), 659.

\bibitem{lt2000} Y.\ T.\ Liu and K.\ S.\ Thorne, Phys.\ Rev.\ D {\bf
  62} (2000), 122002.

\bibitem{sl2001} D.\ H.\ Santamore and Yu.\ Levin, Phys.\ Rev.\ D {\bf
  64}, 042002 (2001).

\bibitem{bc1} A.\ Buonanno and Y.\ Chen, Phys.\ Rev.\ D {\bf 64}
  (2001), 042006.

\bibitem{bc2} A.\ Buonanno and Y.\ Chen, Class.\ Quantum Grav.\ {\bf 18}
  (2001), L95.

\bibitem{ht1998} S.\ A.\ Hughes and K.\ S.\ Thorne, Phys.\ Rev.\ D
  {\bf 58} (1998), 122002.

\bibitem{tw1999} K.\ S.\ Thorne and C.\ J.\ Winstein, Phys.\ Rev.\ D
  {\bf 60} (1999), 082001.

\bibitem{tcreighton} T.\ Creighton, Phys.\ Rev.\ D, submitted;
  gr-qc/0007050.

\bibitem{virgo} F.\ Marion et al., {\it in} Proceedings of the 3rd
  Edoardo Amaldi Conference (S.\ Meshkov, Ed.), AIP Conference
  Proceedings 523, p.\ 110, Melville, New York, 2000.

\bibitem{geo} H.\ L\"uck et al., {\it in} Proceedings of the 3rd
  Edoardo Amaldi Conference (S.\ Meshkov, Ed.), AIP Conference
  Proceedings 523, p.\ 119, Melville, New York, 2000.

\bibitem{tama} M.\ Ando et al., Phys.\ Rev.\ Lett.\ {\bf 86} (2001),
  3950.

\bibitem{noisecomment} The displacement noise ${\tilde x}$ of a
  detector is quoted in units ${\rm cm}/\sqrt{\rm Hz}$ so that the
  squared displacement noise in a band $\Delta f$, $\sigma_{L}^2 =
  \int_{\Delta f} df\,{\tilde x}^2$, has units of ${\rm cm}^2$.
  Likewise, the strain noise ${\tilde h}$ is quoted in units
  $1/\sqrt{\rm Hz}$.

\bibitem{advtama} K.\ Kuroda et al., Int.\ J.\ Mod.\ Phys.\ D {\bf 8}
  (2000), 557.

\bibitem{aciga} D.\ McCleland et al., {\it in} Proceedings of the 3rd
  Edoardo Amaldi Conference (S.\ Meshkov, Ed.), AIP Conference
  Proceedings 523, p.\ 140, Melville, New York, 2000.

\bibitem{whitepaper} E.\ Gustafson, D.\ Shoemaker, K.\ Strain, and R.\
  Weiss, ``LSC White Paper on Detector Research and Development'',
  LIGO Document T990080-00-D (1999).

\bibitem{tuning} S.\ A.\ Hughes, Phys.\ Rev.\ D, in press; also
  gr-qc/0209012.

\bibitem{hulsetaylor} R.\ A.\ Hulse and J.\ H.\ Taylor, Astrophys.\
  J.\ {\bf 195} (1975), L51.

\bibitem{taylorweisberg} J.\ H.\ Taylor and J.\ M.\ Weisberg,
  Astrophys.\ J.\ {\bf 345} (1989), 434.

\bibitem{stairs} I.\ H.\ Stairs, S.\ E.\ Thorsett, J.\ H.\ Taylor, and
  Z.\ Arzoumanian, {\it in} Pulsar Astronomy --- 2000 and Beyond, ASP
  Conference Series Vol.\ 202, ASP, San Francisco, 2000.

\bibitem{bb98} H.\ Bethe and G.\ E.\ Brown, Astrophys.\ J.\ {\bf 506}
  (1998), 780.

\bibitem{pzy98} S.\ F.\ Portegies-Zwart and L.\ R.\ Yungel'son,
  Astron.\ Astrophys.\ {\bf 332} (1998), 173.

\bibitem{fwh99} C.\ L.\ Fryer, S.\ E.\ Woosley, and D.\ H.\ Hartmann,
  Astrophys.\ J.\ {\bf 526} (1999), 152.

\bibitem{pzw2000} S.\ F.\ Portegies-Zwart and S.\ L.\ W.\ McMillan,
  Astrophys.\ J.\ {\bf 528} (2000), L17.

\bibitem{bkb2002} K.\ Belczynski, V.\ Kalogera, and T.\ Bulik,
  Astrophys.\ J.\ {\bf 572} (2002), 407.

\bibitem{nstt97} T.\ Nakamura, M.\ Sasaki, T.\ Tanaka, and K.\ S.\
  Thorne, Astrophys.\ J.\ {\bf 487} (1997), L139.

\bibitem{ictn98} K.\ Ioka, T.\ Chiba, T.\ Tanaka, and T.\ Nakamura,
  Phys.\ Rev.\ D {\bf 58} (1998), 063003.

\bibitem{eardley83} D.\ M.\ Eardley, {\it in} ``Gravitational
  Radiation'' (N.\ Dereulle and T.\ Piran, Eds.), p.\ 257, North
  Holland, Amsterdam, 1983.

\bibitem{chandra69} S.\ Chandrasekhar, ``Ellipsoidal Figures
  of Equilibrium'', Yale University Press, New Haven, 1969.

\bibitem{cent2000} J.\ M.\ Centrella, K.\ C.\ B.\ New, L.\ L.\ Lowe,
  and J.\ D.\ Brown, Astrophys.\ J.\ {\bf 550} (2000), 193.

\bibitem{new2000} K.\ C.\ B.\ New, J.\ M.\ Centrella, and J.\ E.\
  Tohline, Phys.\ Rev.\ D {\bf 62} (2000), 064019.

\bibitem{fhh} C.\ L.\ Fryer, D.\ E.\ Holz, and S.\ A.\ Hughes,
  Astrophys.\ J.\ {\bf 565} (2002), 430.

\bibitem{fw2002} C.\ L.\ Fryer and M.\ S.\ Warren, Astrophys.\ J.\
  {\bf 574} (2002), L65.

\bibitem{jones} D.\ I.\ Jones, Class.\ Quantum Grav.\ {\bf 19}
  (2002), 1255.

\bibitem{ben} B.\ J.\ Owen, {\it in} ``Matters of Gravity'' (J.\
  Pullin, Ed.), Vol.\ 20, p.\ 8; available online at gr-qc/0209085.

\bibitem{cutler02} C.\ Cutler, Phys.\ Rev.\ D, submitted;
  gr-qc/0206051.

\bibitem{lars} L.\ Bildsten, Astrophys.\ J.\ {\bf 501} (1998), L89.

\bibitem{ucb} G.\ Ushomirsky, C.\ Cutler, and L.\ Bildsten, Mon.\
  Not.\ R.\ Astron.\ Soc.\ {\bf 319} (2000), 902.

\bibitem{andersson} N.\ Andersson, Astrophys.\ J.\ {\bf 502} (1998),
  702.

\bibitem{fm98} J.\ L.\ Friedman and S.\ M.\ Morsink, Astrophys.\
  J.\ {\bf 502} (1998), 714.

\bibitem{lom98} L.\ Lindblom, B.\ J.\ Owen, and S.\ M.\ Morsink,
  Phys.\ Rev.\ Lett.\ {\bf 80} (1998), 4843.

\bibitem{nks99} N.\ Andersson, K.\ Kokkotas, and B.\ F.\ Schutz,
  Astrophys.\ J.\ {\bf 510} (1999), 846.

\bibitem{olcsva98} B.\ J.\ Owen, L.\ Lindblom, C.\ Cutler, B.\ F.\
  Schutz, A.\ Vecchio, and N.\ Andersson, Phys.\ Rev.\ D {\bf 58}
  (1998), 084020.

\bibitem{periodic} P.\ R.\ Brady, T.\ Creighton, C.\ Cutler,
  and B.\ F.\ Schutz, Phys.\ Rev.\ D {\bf 57} (1999), 2101.

\bibitem{per_hier} P.\ R.\ Brady and T.\ Creighton, Phys.\ Rev.\
  D {\bf 61} (2000), 082001.

\bibitem{rls2000} L.\ Rezzolla, F.\ K.\ Lamb, and S.\ L.\ Shapiro,
  Astrophys.\ J.\ {\bf 531} (2000), 139.

\bibitem{lo2002} L.\ Lindblom and B.\ J.\ Owen, Phys.\ Rev.\ D, {\bf
  65} (2002), 063006.

\bibitem{arrasetal} P.\ Arras, E.\ E.\ Flanagan, S.\ M.\ Morsink, A.\
  K.\ Schenk, S.\ A.\ Teukolsky, and I.\ Wasserman, Phys.\ Rev.\ D,
  submitted; astro-ph/0202345.

\bibitem{gressmanetal} P.\ Gressman, L.-M.\ Lin, W.-M.\ Suen, N.\
  Stergioulas, and J.\ L.\ Friedman, Phys.\ Rev.\ D {\bf 66} (2002),
  041303.

\bibitem{heyl} J.\ S.\ Heyl, Astrophys.\ J.\ {\bf 574} (2002, L57.

\bibitem{fa2002} J.\ L.\ Friedman and N.\ Andersson {\it in} ``Matters
  of Gravity'' (J.\ Pullin, Ed.), Vol.\ 20, p.\ 5; available online at
  gr-qc/0209085.

\bibitem{turner1} M.\ S.\ Turner, Phys.\ Rev.\ D {\bf 55} (1997), 45.

\bibitem{bruce96} B.\ Allen, {\it in} ``Proceedings of the Les Houches
  School on Astrophysical Sources of Gravitational Waves'' (J.-A.\
  Marck and J.-P.\ Lasota, Eds.), Cambridge University Press,
  Cambridge, 1996; also gr-qc/9604033.

\bibitem{kmk2001} A.\ Kosowsky, A.\ Mack, and T.\ Kahniashvili, {\it
  in} ``Astrophysical sources for ground-based gravitational wave
  detectors'', AIP Conference Proceedings Vol.\ 575, p.\ 191, American
  Institute of Physics, Melville, NY, 2001.

\bibitem{hogan_prl} C.\ J.\ Hogan, Phys.\ Rev.\ Lett.\ {\bf 85}
  (2000), 2044.

\bibitem{hogan_prd} C.\ J.\ Hogan, Phys.\ Rev.\ D {\bf 62} (2000),
  121302.

\bibitem{rs1} L.\ Randall and R.\ Sundrum, Phys.\ Rev.\ Lett.\ {\bf
  83}, 3370 (1999).

\bibitem{rs2} L.\ Randall and R.\ Sundrum, Phys.\ Rev.\ Lett.\ {\bf
  83}, 4690 (1999).

\bibitem{sss2002} V.\ Sahni, M.\ Sami, and T.\ Souradeep, Phys.\ Rev.\
  D {\bf 65} (2002), 023518.

\bibitem{maggiore} M.\ Maggiore, Phys.\ Rep.\ {\bf 331} (2000), 283.

\bibitem{ar} B.\ Allen and J.\ D.\ Romano, Phys.\ Rev.\ D {\bf 59}
  (1999), 102001.

\bibitem{prephaseA} K.\ Danzmann, Ed., LISA Pre-Phase A Report, 2nd
  Edition, Report MPQ-233, Max-Planck Institut f\"ur Quantenoptik,
  Garching, Germany, 1998.

\bibitem{lisa1} Proceedings of the First International LISA Symposium:
  Class.\ Quantum Grav.\ {\bf 14} (1997), 1397 -- 1585.

\bibitem{lisa2} W.\ M.\ Folkner, Ed., Proceedings of the Second
  International LISA Symposium, American Institute of Physics,
  Woodbury, N.\ Y., 1998.

\bibitem{jointproposal} P.\ L.\ Bender, {\it in} ``Gravitational
  Waves'' (I.\ Ciufolini, V.\ Gorini, V.\ Moschella, and P.\ Fre,
  Eds.) Institute of Physics Publishing, Bristol, UK, 2001, p.\ 115.

\bibitem{lhh} S.\ L.\ Larson, W.\ A.\ Hiscock, and R.\ W.\ Hellings,
  Phys.\ Rev.\ D {\bf 62} (2000), 062001.

\bibitem{cutler98} C.\ Cutler, Phys.\ Rev.\ D {\bf 57} (1998), 7089.

\bibitem{mm2001} M.\ Milosavljevi\'c and D.\ Merritt, Astrophys.\
  J.\ {\bf 563} (2001), 34.

\bibitem{av2002} P.\ J.\ Armitage and P.\ Natarajan, Astrophys.\ J.\
  {\bf 567} (2002), L9.

\bibitem{me2002} D.\ Merritt and R.\ Ekers, Science {\bf 297} (2002),
  1310.

\bibitem{mhn2001} K.\ Menou, Z.\ Haiman, and V.\ K.\ Narayanan,
  Astrophys.\ J.\ {\bf 558} (2001), 535.

\bibitem{sigurdssonrees} S.\ Sigurdsson and M.\ J.\ Rees,
  Mon.\ Not.\ R.\ Astron.\ Soc.\ {\bf 284} (1997), 318.

\bibitem{sigurdsson} S.\ Sigurdsson, Class.\ Quantum Grav.\ {\bf 14}
  (1997), 1425.

\bibitem{aet99} J.\ W.\ Armstrong, F.\ B.\ Estabrook, and M.\ Tinto,
  Astrophys.\ J.\ {\bf 527} (1999), 814.

\bibitem{hb2001} C.\ J.\ Hogan and P.\ L.\ Bender, Phys.\ Rev.\ D
  {\bf 64} (2001), 062002.

\bibitem{turner2} M.\ S.\ Turner, Proc.\ Astron.\ Soc.\ Pacific {\bf
  111} (1999), 264.

\bibitem{kamkos99} M.\ Kamionkowski and A.\ Kosowsky, Ann.\ Rev.\
  Nucl.\ Part.\ Sci.\ {\bf 49} (1999), 77.

\bibitem{kamkos98} M.\ Kamionkowski and A.\ Kosowsky, Phys.\ Rev.\ D
  {\bf 57} (1998), 685.

\bibitem{kks97} M.\ Kamionkowski, A.\ Kosowsky, and A.\ Stebbins,
  Phys.\ Rev.\ D {\bf 55} (1997), 7368.

\bibitem{emodes} J.\ Kovac et al., Astrophys.\ J., submitted;
  astro-ph/0209478.

\bibitem{hhz2002} W.\ Hu, M.\ M.\ Hedman, and M.\ Zaldarriaga, Phys.\
  Rev.\ D, submitted; astro-ph/0210096.

\bibitem{jkw2000} A.\ H.\ Jaffe, M.\ Kamionkowski, and L.\ Wang,
  Phys.\ Rev.\ D {\bf 61} (2000), 083501.

\bibitem{tetal2000} M.\ Tegmark, D.\ J.\ Eisenstein, W.\ Hu, and A.\
  de Oliveira-Costa, Astrophys.\ J.\ {\bf 530} (2000), 133.

\bibitem{psb2000} S.\ Prunet, S.\ K.\ Sethi, and F.\ R.\ Bouchet,
  Mon.\ Not.\ R.\ Astron.\ Soc.\ {\bf 314} (2000), 348.

\bibitem{baccietal2002} C.\ Baccigalupi et al., Mon.\ Not.\ R.\
  Astron.\ Soc., submitted; astro-ph/0209591.

\bibitem{kck2002} M.\ Kesden, A.\ Cooray, and M.\ Kamionkowski, Phys.\
  Rev.\ Lett.\ {\bf 89} (2002), 1304.

\bibitem{zs97} M.\ Zaldarriaga and U.\ Seljak, Phys.\ Rev.\ D {\bf 55}
  (1997), 1830.

\bibitem{thorne80} K.\ S.\ Thorne, Rev.\ Mod.\ Phys.\ {\bf 52} (1980,
  299.

\bibitem{sazhin} M.\ V.\ Sazhin, Soviet Astronomy {\bf 22} (1978), 36.

\bibitem{det79} S.\ Detweiler, Astrophys.\ J.\ {\bf 234} (1979), 1100.

\bibitem{lommen2002} A.\ N.\ Lommen, {\it to appear in} ``Proceedings
  of the 270 WE-Heraeus Seminar on Neutron Stars, Pulsars and
  Supernova Remnants, Jan. 21-25, 2002, Physikzentrum Bad Honnef''
  (W.\ Becker, H.\ Lesch, and J.\ Truemper, Eds.); also
  astro-ph/0208572.

\bibitem{ktr94} V.\ M.\ Kaspi, J.\ H.\ Taylor, and M.\ F.\ Ryba,
  Astrophys.\ J.\ {\bf 428} (1994), 713.

\bibitem{jb2002} A.\ Jaffe and D.\ C.\ Backer, Astrophys.\ J., in
  press; also astro-ph/0210148.

\bibitem{zcm} X.\ Zhuge, J.\ M.\ Centrella, and S.\ L.\ W.\ McMillan,
  Phys.\ Rev.\ D {\bf 50} (1996), 7261.

\bibitem{rs} F.\ A.\ Rasio and S.\ L.\ Shapiro, Class.\ Quantum Grav.\
  {\bf 16} (1999), R1.

\bibitem{fgrt} J.\ A.\ Faber, P.\ Grandcl\'ement, F.\ A.\ Rasio, and
  K.\ Taniguchi, Phys.\ Rev.\ Lett., in press; also astro-ph/0204397.

\bibitem{shibata} M.\ Shibata, Prog.\ Theor.\ Phys.\ {\bf 96} (1996),
  917.

\bibitem{wigginslai} P.\ Wiggins and D.\ Lai, Astrophys.\ J.\ {\bf 532}
  (2000), 530.

\bibitem{wald} R.\ M.\ Wald, ``General Relativity'', University of
  Chicago Press, Chicago, 1984.

\bibitem{fh1998} E.\ E.\ Flanagan and S.\ A.\ Hughes, Phys.\ Rev.\ D
  {\bf 57} (1998), 4535.

\bibitem{kerr} R.\ P.\ Kerr, Phys.\ Rev.\ Lett.\ {\bf 11} (1962), 237.

\bibitem{nohair} By ``no-hair theorem'', we refer to a collection of
  works which establish that the Kerr solution is the only rotating,
  stationary black hole solution {\cite{carter,robinson}}, and that
  radiation emission always and quickly drives a distorted object to
  the Kerr solution {\cite{price1,price2}}. References
  {\cite{carter,robinson}} generalize earlier work by Werner Israel
  {\cite{israel}} proving that the Schwarzschild solution is the
  unique solution describing a non-rotating black hole.  We ignore the
  possibility of charged black holes, which are astrophysically
  irrelevant (they are quickly neutralized in any astrophysical
  environment by interstellar plasma).

\bibitem{carter} B.\ Carter, Phys.\ Rev.\ Lett.\ {\bf 26} (1971), 331.

\bibitem{robinson} D.\ C.\ Robinson, Phys.\ Rev.\ Lett.\ {\bf 34}
  (1975), 905.

\bibitem{israel} W.\ Israel, Phys.\ Rev.\ {\bf 164} (1967), 1776.

\bibitem{price1} R.\ H.\ Price, Phys.\ Rev.\ D {\bf 5} (1972), 2419.

\bibitem{price2} R.\ H.\ Price, Phys.\ Rev.\ D {\bf 5} (1972), 2439.

\bibitem{thud} This fact was first pointed out to me by Sam Finn.

\bibitem{teuk73} S.\ A.\ Teukolsky, Astrophys.\ J.\ {\bf 185} (1973),
  635.

\bibitem{leaver} E.\ W.\ Leaver, Proc.\ R.\ Soc.\ Lond.\ {\bf A402}
  (1985), 285.

\bibitem{echeverria} F.\ Echeverria, Phys.\ Rev.\ D {\bf 40} (1989),
  3194.

\bibitem{lehner} L.\ Lehner, Class.\ Quantum Grav.\ {\bf 18} (2001),
  R25.

\bibitem{kst2001} L.\ E.\ Kidder, M.\ A.\ Scheel, and S.\ A.\
  Teukolsky, Phys.\ Rev.\ D {\bf 64} (2001), 064017.

\bibitem{ls2002} L.\ Lindblom and M.\ A.\ Scheel, Phys.\ Rev.\ D,
  in press; also gr-qc/0206035.

\bibitem{sklpt} M.\ A.\ Scheel, L.\ E.\ Kidder, L.\ Lindblom, H.\ P.\
  Pfeiffer, and S.\ A.\ Teukolsky, Phys.\ Rev.\ D, submitted; also
  gr-qc/0209115.

\bibitem{bcc} J.\ Baker, M.\ Campanelli, and C.\ Lousto,
  Phys.\ Rev.\ D {\bf 65} (2002), 044001.

\bibitem{bclt} J.\ Baker, M.\ Campanelli, C.\ O.\ Lousto, and R.\
  Takahashi, Phys.\ Rev.\ D {\bf 65} (2002), 124012.

\bibitem{wmm} J.\ R.\ Wilson, G.\ J.\ Mathews, and P.\ Marronetti,
  Phys.\ Rev.\ D {\bf 54} (1996), 1317.

\bibitem{bcsst} T.\ W.\ Baumgarte, G.\ B.\ Cook, M.\ A.\ Scheel,
  S.\ L.\ Shapiro, and S.\ A.\ Teukolsky, Phys.\ Rev.\ Lett.\ {\bf 79}
  (1997), 1182.

\bibitem{bgm} S.\ Bonazzola, E.\ Gourgoulhon, and J.-A.\ Marck,
  Phys.\ Rev.\ D {\bf 56} (1997), 7740.

\bibitem{shibata98} M.\ Shibata, Phys.\ Rev.\ D {\bf 58} (1998),
  024012.

\bibitem{ue2000} K.\ and Y.\ Eriguchi, Phys.\ Rev.\ D {\bf 61} (2000),
  124023.

\bibitem{ggb1} E.\ Gourgoulhon, P.\ Grandcl\'ement, and S.\ Bonazzola,
  Phys.\ Rev.\ D {\bf 65 }(2002), 044020.

\bibitem{ggb2} E.\ Gourgoulhon, P.\ Grandcl\'ement, and S.\ Bonazzola,
  Phys.\ Rev.\ D {\bf 65 }(2002), 044021.

\bibitem{bd99} A.\ Buonanno and T.\ Damour, Phys.\ Rev.\ D {\bf 59}
  (1999), 084006.

\bibitem{bd2000} A.\ Buonanno and T.\ Damour, Phys.\ Rev.\ D {\bf 62}
  (2000), 064015.

\bibitem{djs2000} T.\ Damour, P.\ Jaranowski, and G.\ Sch\"afer,
  Phys.\ Rev.\ D {\bf 62} (2000), 084011.

\bibitem{d2001} T.\ Damour, Phys.\ Rev.\ D {\bf 64} (2001), 124013.

\bibitem{dgg} T.\ Damour, E.\ Gourgoulhon, and P.\ Grandcl\'ement,
  Phys.\ Rev.\ D {\bf 66} (2002), 024007.

\bibitem{bbr1980} M.\ C.\ Begelman, R.\ D.\ Blandford, M.\ J.\ Rees,
  Nature {\bf 287} (1980), 307.

\bibitem{fc93} L.\ S.\ Finn and D.\ F.\ Chernoff, Phys.\ Rev.\ D
  {\bf 47} (1993), 2198.

\bibitem{cf94} C.\ Cutler and E.\ E.\ Flanagan, Phys.\ Rev.\ D {\bf
  49} (1994), 2658.

\bibitem{pw95} E.\ Poisson and C.\ M.\ Will, Phys.\ Rev.\ D {\bf 52}
  (1995), 848.

\bibitem{untangle} S.\ A.\ Hughes, Mon.\ Not.\ R.\ Astron.\ Soc.\
  {\bf 331} (2002), 805.

\bibitem{BBHcandle} D.\ E.\ Holz and S.\ A.\ Hughes, in preparation.

\bibitem{bertrand} In fact, it is simple to show that closed orbits in
  a central potential $V \propto r^n$ occur only for $n = -1$ and $n =
  2$; this is known as Bertrand's theorem.  See H.\ Goldstein,
  ``Classical Mechanics'', 2nd ed., Sec.\ 3-6 and Appendix A,
  Addison-Wesley Publishing Co., Reading, Massachusetts, 1980.

\bibitem{grace} Information about the GRACE mission can be found at
  the WWW URL\\ {\tt http://www.csr.utexas.edu/grace/overview.html}.

\bibitem{phystoday} B.\ G.\ Levi, Physics Today {\bf 55} (2002), 17.

\bibitem{csw86} M.\ Colpi, S.\ L.\ Shapiro, and I.\ Wasserman,
  Phys.\ Rev.\ Lett.\ {\bf 57} (1986), 2485.

\bibitem{lm92} A.\ R.\ Liddle and M.\ S.\ Madsen, Int.\ J.\ Mod.\
  Phys.\ D {\bf 1} (1992), 101.

\bibitem{jetzer92} P.\ Jetzer, Phys.\ Rep.\ {\bf 220} (1992), 163.

\bibitem{lp92} T.\ D.\ Lee and Y.\ Pang, Phys.\ Rep.\ {\bf 221}
  (1992), 251.

\bibitem{ds2000} M.\ P.\ Dabrowski and F.\ E.\ Schnuck, Astrophys.\
  J.\ {\bf 535} (2000), 316.

\bibitem{fintan} F.\ Ryan, Phys.\ Rev.\ D {\bf 56} (1997), 1845.

\bibitem{wilms} J.\ Wilms, C.\ S.\ Reynolds, M.\ C.\ Begelman, J.\
  Reeves, S.\ Molendi, R.\ Staubert, and E.\ Kendziorra, Mon.\ Not.\
  R.\ Astro.\ Soc.\ {\bf 328} (2001), L27.

\bibitem{elvis} M.\ Elvis, G.\ Risaliti, and G.\ Zamorani, Astrophys.\
  J.\ {\bf 519} (2002), 89.

\bibitem{mtw} C.\ W.\ Misner, K.\ S.\ Thorne, and J.\ A.\ Wheeler,
  ``Gravitation'', Freeman, San Francisco, 1973.

\bibitem{bt87} J.\ Binney and S.\ Tremain, ``Galactic Dynamics'',
  Princeton University Press, Princeton, 1987.

\bibitem{poisson93} E.\ Poisson, Phys.\ Rev.\ D {\bf 47} (1993), 1497.

\bibitem{cfps} C.\ Cutler, L.\ S.\ Finn, E.\ Poisson, and G.\ J.\
  Sussman, Phys.\ Rev.\ D {\bf 47} (1993), 1511.

\bibitem{akop} T.\ Apostolatos, D.\ Kennefick, A.\ Ori, and E.\
  Poisson, Phys.\ Rev.\ D {\bf 47} (1993), 5376.

\bibitem{ckp} C.\ Cutler, D.\ Kennefick, and E.\ Poisson, Phys.\ Rev.\
  D {\bf 50} (1994), 3816.

\bibitem{dank98} D.\ Kennefick, Phys.\ Rev.\ D {\bf 58} (1998),
  064012.

\bibitem{paperI} S.\ A.\ Hughes, Phys.\ Rev.\ D {\bf 61} (2000),
  084004.

\bibitem{paperII} S.\ A.\ Hughes, Phys.\ Rev.\ D {\bf 64} (2001),
  064004.

\bibitem{zoomwhirl} K.\ Glampedakis and D.\ Kennefick, Phys.\ Rev.\ D
  {\bf 66} (2002), 044002.

\bibitem{mst97} Y.\ Mino, M.\ Sasaki, and T.\ Tanaka, Phys.\ Rev.\ D
  {\bf 55} (1997), 3457.

\bibitem{qw97} T.\ C.\ Quinn and R.\ M.\ Wald, Phys.\ Rev.\ D {\bf 56}
  (1997), 3381.

\bibitem{wiseman2000} A.\ G.\ Wiseman, Phys.\ Rev.\ D {\bf 61} (2000),
  084014.

\bibitem{bo2000} L.\ Barack and A.\ Ori, Phys.\ Rev.\ D {\bf 61}
  (2000), 082001.

\bibitem{burko2000} L.\ M.\ Burko, Phys.\ Rev.\ Lett.\ {\bf 84}
  (2000), 4529.

\bibitem{lousto2000} C.\ Lousto, Phys.\ Rev.\ Lett.\ {\bf 84} (2000),
  5251.

\bibitem{bmnos} L.\ Barack, Y.\ Mino, H.\ Nakano, A.\ Ori and M.\
  Sasaki, Phys.\ Rev.\ Lett.\ {\bf 88} (2002), 091101.

\bibitem{pp2002} M.\ J.\ Pfenning and E.\ Poisson, Phys.\ Rev.\ D {\bf
  65} (2002), 084001.

\bibitem{ko96} D.\ Kennefick and A.\ Ori, Phys.\ Rev.\ D {\bf 53}
  (1996), 4319.

\bibitem{ryan96} F.\ D.\ Ryan, Phys.\ Rev.\ D {\bf 53} (1996), 3064.

\bibitem{mino_thesis} Y.\ Mino, unpublished Ph.\ D.\ thesis, Kyoto
  University, 1996.

\bibitem{membrane} K.\ S.\ Thorne, R.\ H.\ Price, and D.\ A.\
  MacDonald, ``Black Holes: The Membrane Paradigm'', Yale University
  Press, New Haven, 1986.

\bibitem{zw_curt} This name was originally coined by Curt Cutler to
  describe the nature of eccentric inspirals into non-rotating black
  holes; as Glampedakis and Kennefick show {\cite{zoomwhirl}}, this
  term is even more appropriate to describe inspirals into rotating
  holes.

\end{thebibliography}
\end{document}